\documentclass[runningheads]{llncs}
\usepackage[utf8]{inputenc}

\usepackage{amsmath}
\usepackage{amssymb} 
\usepackage{graphicx} 
\usepackage{bussproofs} 
\usepackage{soul} 
\usepackage[titlenumbered,ruled]{algorithm2e} 
\usepackage{color} 
\usepackage{hyperref} 
\usepackage[most]{tcolorbox} 
\usepackage{tikz} 
\usetikzlibrary{matrix} 

\newcommand{\cmatrix}[2]{
  \ensuremath{\left [ \rotatebox[origin=c]{90}{\reflectbox{ \begin{tikzpicture}[baseline=-1.5ex]
    \matrix (m) [matrix of math nodes, ampersand replacement=\&, nodes={rotate=90,xscale=-1}] { #1 };
    #2
  \end{tikzpicture} }} \right ]}
}

\newcommand{\defined}{\:\raisebox{1ex}{\scalebox{0.5}{\ensuremath{\mathrm{def}}}}\hskip-1.65ex\raisebox{-0.1ex}{\ensuremath{=}}\:}
\newcommand{\tuple}[1]{\ensuremath{\langle #1 \rangle}}
\newcommand{\II}{\ensuremath{\mathcal{I}}}
\newcommand{\OO}{\ensuremath{\mathcal{O}}}
\newcommand{\sats}{\Vdash}
\newcommand{\ou}{\sqcup}
\newcommand{\e}{\sqcap} 
\newcommand{\pf}{\bullet}
\newcommand{\nao}{\neg}
\newcommand{\sub}{\sqsubseteq}
\newcommand{\alc}{\ensuremath{\mathcal{ALC}}}
\newcommand{\alcb}{\ensuremath{\mathcal{ALC}^\pf}}
\newcommand{\alch}{\ensuremath{\mathcal{ALCH}}}
\newcommand{\alchb}{\ensuremath{\mathcal{ALCH}^\pf}}
\newcommand{\alctcm}{\ensuremath{\mathcal{ALC}-\theta~CM}}
\newcommand{\CN}{\ensuremath{\mathsf{C}}} 
\newcommand{\RN}{\ensuremath{\mathsf{R}}} 
\newcommand{\IN}{\ensuremath{\mathsf{I}}} 
\newcommand{\KB}{\mathcal{KB}}
\newcommand{\K}{\mathcal{K}}
\newcommand{\T}{\mathcal{T}}
\newcommand{\R}{\mathcal{R}}
\newcommand{\A}{\mathcal{A}}
\newcommand{\MuggleF}{\ensuremath{\mathsf{Muggle}}}
\newcommand{\WizardF}{\ensuremath{\mathsf{Wizard}}}
\newcommand{\SpellF}{\ensuremath{\mathsf{Spell}}}
\newcommand{\castsF}{\ensuremath{\mathsf{casts}}}
\newcommand{\hermioneF}{\ensuremath{\mathsf{hermione}}}
\newcommand{\HalfBloodWizardF}{\ensuremath{\mathsf{HalfBloodWizard}}}
\newcommand{\WandF}{\ensuremath{\mathsf{Wand}}}
\newcommand{\attachedF}{\ensuremath{\mathsf{attachedWith}}}
\newcommand{\masterF}{\ensuremath{\mathsf{masterOf}}}
\newcommand{\voldemortF}{\ensuremath{\mathsf{lordvoldemort}}}
\newcommand{\elderWandF}{\ensuremath{\mathsf{elderwand}}}

\newcommand{\ronF}{\ensuremath{\mathsf{ronweasley}}}
\newcommand{\datesF}{\ensuremath{\mathsf{marriedTo}}}
\newcommand{\knowsF}{\ensuremath{\mathsf{hasPartner}}}
\newcommand{\PBWizardF}{\ensuremath{\mathsf{PureBloodWizard}}}

\newcommand{\true}{\ensuremath{\mathsf{true}}}
\newcommand{\false}{\ensuremath{\mathsf{false}}}

\makeatletter
\providecommand{\bigsqcap}{%
  \mathop{%
    \mathpalette\@updown\bigsqcup
  }%
}
\newcommand*{\@updown}[2]{%
  \rotatebox[origin=c]{180}{$\m@th#1#2$}%
}
\makeatother

\SetKwComment{Comment}{/* }{ */}
\SetKwInOut{Input}{Input}
\SetKwInOut{Output}{Output}
\SetKwProg{Proof}{Proof}{}{end}
\SetKwProg{Inconsistent}{Inconsistent}{}{end}

\newcommand{\df}[1]{\ul{#1}}

\makeatletter
\def\moverlay{\mathpalette\mov@rlay}
\def\mov@rlay#1#2{\leavevmode\vtop{%
   \baselineskip\z@skip \lineskiplimit-\maxdimen
   \ialign{\hfil$\m@th#1##$\hfil\cr#2\crcr}}}
\newcommand{\charfusion}[3][\mathord]{
    #1{\ifx#1\mathop\vphantom{#2}\fi
        \mathpalette\mov@rlay{#2\cr#3}
      }
    \ifx#1\mathop\expandafter\displaylimits\fi}
\makeatother

\newcommand\smallmath[2]{#1{\raisebox{\dimexpr \fontdimen 22 \textfont 2
      - \fontdimen 22 \scriptscriptfont 2 \relax}{$\scriptscriptstyle #2$}}}
\newcommand\smalltimes{\smallmath\mathbin\times}

\newcommand{\clup}{\charfusion[\mathbin]{\cup}{\smalltimes}}

\newtcolorbox{reviewer}{
enhanced,
fonttitle=\sffamily\bfseries,
fontupper=\small\sffamily,
coltitle=orange,
title={\scriptsize Reviewer:},
boxrule=0pt,
frame hidden,
borderline west={4pt}{0pt}{orange},
colback=yellow!12!white,
colbacktitle=yellow!12!white
}

\begin{document}

\title{A connection method for a defeasible extension of \alch}
%
%
\author{Renan Fernandes\inst{1}\orcidID{0000-0001-9553-5515} \and
Fred Freitas\inst{1}\orcidID{0000-0003-0425-6786}
\and 
Ivan Varzinczak\inst{2,3,4}\orcidID{0000-0002-0025-9632}
\and 
Pedro PM Farias\inst{1,5}\orcidID{0000-0002-6344-4448}
}
\authorrunning{R. Fernandes et al.}
%
\institute{Centro de Informática, Universidade Federal de Pernambuco, Brazil \\
\email{\{rlf5,fred,ppmf\}@cin.ufpe.br}
\and
LIASD, Université Paris~8, France\\
\and
CAIR, University of Cape Town, South Africa \\
\and
ISTI--CNR, Italy\\
\email{ivan.varzinczak@univ-paris8.fr}
\and
ARCE, Public Services Regulation Agency-CE, Brazil
}
\maketitle

\begin{abstract}
This paper proposes a connection method à la Bibel for an exception-tolerant family of description logics (DLs). As for the language, we assume the DL~\alch\ extended with two typicality operators: one on (complex) concepts and one on role names. The language is a variant of defeasible DLs, as broadly studied in the literature over the past decade, in which most of these can be embedded. We revisit the definition of the matrix representation of a knowledge base and establish the conditions for a given axiom to be provable. 
We show that the algorithm terminates, is sound and complete w.r.t. a DL version of the preferential semantics widely adopted in non-monotonic reasoning.

\end{abstract}

\section{Introduction}

The problem of modelling exceptions in ontologies and reasoning meaningfully in their presence has received a great deal of attention over the past decade. Among the emblematic approaches put forward in the literature feature Giordano et al.'s description logics of typicality~\cite{GiordanoEtAl2007,GiordanoEtAl2009,GiordanoEtAl2015}, Britz et al.'s defeasible subsumption relations~\cite{BritzEtAl2008,BritzEtAl2021}, Bonatti et al.'s light-weight DLs of normality~\cite{BonattiEtAl2011,BonattiEtAl2015,BonattiSauro2017}, Bozzato~et al.'s reduction to Datalog~\cite{BozzatoEtAl2018}, besides Casini and Straccia's seminal work on the computational counterpart of non-monotonic entailment in DLs~\cite{CasiniStraccia2010,CasiniStraccia2013} along with its implementation~\cite{CasiniEtAl2015}. These investigations have given rise to a whole family of defeasible description logics of varying expressive power and with the ability to handle exceptions at both the \textit{modelling} and the \textit{reasoning} levels in a number of ways~\cite{Bonatti2019,BritzVarzinczak2017-DL,CasiniEtAl2014,CasiniEtAl2019,PenselTurhan2018}.

One of the interesting characteristics of some of the aforementioned approaches is the fact that depending on the underlying DL that is assumed and given certain conditions on how exceptionality (or typicality) is expressed, the kind of non-monotonic reasoning that is performed can be reduced to (a polynomial number of calls to) \textit{classical} entailment check. Therefore, the study of automated deduction for the various flavours of defeasible~DLs and its potential reduction to classical reasoning remains a relevant and active research topic in logic-based artificial intelligence.

The development of proof methods for defeasible description logics has, in a sense, followed those for classical~DLs. As a result, the overwhelming majority of existing decision procedures for reasoning with defeasible ontologies are based on semantic tableaux~\cite{BritzVarzinczak2017-DL,BritzVarzinczak2019-TABLEAUX,GiordanoEtAl2009,GiordanoEtAl2013,Varzinczak2018}. Notwithstanding the commonly extolled virtues of tableau systems, there are equally viable alternatives in the literature on automated theorem proving (ATP). One prominent example is the connection method (CM)~\cite{bibel:cm}, initially defined by W.~Bibel in the late '70s, which earned a good reputation in the field of ATP in the '80s and '90s. In particular, the connection method has recently been revived in the context of (classical) modal and description logics~\cite{jens:modal,jens:modal2022,fred:calculus,fred:card.calculus}.

The connection method consists in a \textit{direct} proof procedure of which the main internal structure is a matrix representation of the knowledge base and associated query. It lends itself to a more parsimonious usage of memory during proof searches. Indeed, contrary to tableaux and resolution, the connection method does not create intermediate clauses or sentences along the way, keeping its search space confined to the boundaries of the matrix it started off with. The first connection calculus for (classical) description logics, \alctcm~\cite{fred:calculus}, incorporates several features of most~DL proof systems, such as blocking, as well as some useful simplifications so that no variables, Skolem functions or unification are needed. Moreover, a C++ implementation of the method, \textsc{Raccoon}~\cite{dimas:raccoon}, has been developed.\footnote{\url{https://github.com/dmfilho/raccoon}} Worthy of mention is the fact that, despite incorporating \textit{none} of the optimisations commonly done for DL tableaux systems, \textsc{Raccoon} performed competitively in reasoning over~\alc\ ontologies when compared to cutting-edge highly-optimised tableau-based reasoners which had ranked high in  DL reasoners' past competitions.\footnote{See \url{https://goo.gl/V9Ewkv} for details on the comparison.}


In this paper, we provide a concrete method for a defeasible~DL at least as expressive as those frequently considered in the literature. We hope our constructions will serve as a springboard for developing connection methods for defeasible DLs of varying expressive power and further non-monotonic reasoning extensions of \textsc{Raccoon}.


\section{The defeasible DL~\alchb}\label{Preliminaries}

The defeasible DL $\alchb$\cite{Varzinczak2018} is an extension of~$\alch$\cite{BaaderEtAl2017} with typicality operators on both complex concepts and role names. Intuitively, a concept expression of the form~$\pf D$ denotes the most typical (alias normal) objects in the class~$D$, whereas a role expression of the form~$\pf r$, with~$r$ an atomic role denotes the most typical instances of the relationship represented by~$r$. To give a glimpse of~$\alchb$'s expressive power, the axioms
\begin{center}
$\pf\MuggleF\sub\lnot\WizardF$, \\
$\HalfBloodWizardF\sub\exists\castsF.\SpellF\e\lnot{\pf}\WizardF$, and \\
$\pf\WizardF\sub\exists{\pf}\attachedF.\WandF$
\end{center}

\noindent specify, respectively, that ``typical muggles are not wizards'', ``half-blood wizards cast spells but are not typical wizards'', and ``typical wizards have a typical attachment with a wand''. The RBox axiom $\masterF\sub{\pf}\attachedF$ states that ``to be a master of (a wand) means to be typically attached with (that wand)''. Furthermore, the ABox assertion $\lnot{\pf}\attachedF(\voldemortF,\elderWandF)$ formalises the intuition that $\voldemortF$ is not typically attached with $\elderWandF$.

We assume finite sets of concepts, roles, and individual names, denoted resp., with \CN, \RN, and \IN. We denote atomic concepts with $A, B,\ldots$, with $r, s,\ldots$ role names, and $a,b,\ldots$ individual names. Complex roles of~\alchb\ are denoted with~$R,S,\ldots$ and defined by the rule: \[
R ::= \RN\mid \pf R
\]

Complex concepts of \alchb\ are denoted with $D,E,\ldots$ and are built according to the following grammar:
\[
D ::= \CN \mid \top \mid \bot \mid \nao{D} \mid D \e D \mid D \ou D \mid \forall R.D \mid \exists R.D \mid \pf D
\]

The definitions of axiom, GCI, assertion, TBox, RBox (allowing for role subsumption axioms of the form $R\sub S$) and ABox are as in the classical \alch\ case. If $\T$, $\R$  and $\A$ are, respectively, a TBox, an RBox and an ABox, with $\K=(\T,\R,\A)$, we denote henceforth a \textit{knowledge base} (alias ontology), frequently abbreviated as~KB. 

\begin{example}[Wizarding-World Scenario] \label{ex:pref}
Assume we are interested in modelling facts about the wizarding-world and its wonderful features. We have the atomic concepts $\CN = \{\MuggleF, \WizardF, \PBWizardF\}$ representing, respectively, the class of muggles, wizards, and pure-blood wizards. As for the set of atomic roles, we have $\RN = \{\datesF, \knowsF\}$, representing a marriage and a partnership between two people (wizards or muggles). The set of individuals $\IN$ is $\{\hermioneF,\ronF\}$. Below is an example of an \alchb\ knowledge base $\K = (\T,\R,\A)$ for the wizarding-world scenario.
\[
    \T = \begin{Bmatrix}
        \pf \MuggleF \sub \nao{\WizardF}, \\
        \PBWizardF \sub \forall \knowsF . \WizardF
    \end{Bmatrix}
\]
\[
    \R = \begin{Bmatrix}
        \datesF \sub \pf \knowsF
    \end{Bmatrix}
\]
\[
    \quad\quad \A = \begin{Bmatrix}
        \MuggleF (\hermioneF), \\
        \PBWizardF (\ronF), \\
        \datesF (\ronF, \hermioneF)
    \end{Bmatrix}
\]
\end{example}

The semantics of \alchb\ extends that of classical \alch\ and is in terms of partially-ordered structures called bi-ordered interpretations. Before introducing these, we recall a few notions.

A binary relation is a \textit{strict partial order} if it is irreflexive and transitive. If~$<$ is a strict partial order on a given set~$X$, with $\min_{<}X\defined\{x\in X \mid $ there is no $y\in X$ s.t. $y<x\}$ we denote the \textit{minimal elements} of~$X$ w.r.t.~$<$. A strict partial order on a set~$X$ is \textit{well-founded} if, for every~$\emptyset\neq X'\subseteq X$, we have $\min_{<}X'\neq\emptyset$.

\begin{definition}[Bi-ordered interpretation]
An \alchb\ \df{bi-ordered~interpretation} is a tuple $\OO\defined\tuple{\Delta^{\OO},\cdot^{\OO},<^{\OO},\ll^{\OO}}$ s.t.  $\tuple{\Delta^{\OO}, \cdot^{\OO}}$ is a classical~\alch\ interpretation, and both $<^{\OO}\subseteq\Delta^{\OO}\times\Delta^{\OO}$ and $\ll^{\OO}\subseteq(\Delta^{\OO}\times\Delta^{\OO})\times(\Delta^{\OO}\times\Delta^{\OO})$ are well-founded strict partial orders.
\end{definition}

Given $\OO=\tuple{\Delta^{\OO},\cdot^{\OO},<^{\OO},\ll^{\OO}}$, the intuition of $\Delta^{\OO}$ and $\cdot^{\OO}$ is the same as in a standard~\alch\ interpretation. The intuition underlying the orderings~$<^{\OO}$ and~$\ll^{\OO}$ is that they play the role of \textit{preference relations} (or \textit{normality} orderings): the objects (resp.~pairs) that are lower down in the ordering~$<^{\OO}$ (resp.~$\ll^{\OO}$) are deemed more normal (or typical) than those higher up in~$<^{\OO}$ (resp.~$\ll^{\OO}$). Within the context of (the interpretation of) a concept~$D$ (resp.\ role~$R$), $<^{\OO}$ (resp.~$\ll^{\OO}$) therefore allows us to single out the most normal representatives falling under~$D$ (resp.~$R$), which is the intuition of the semantics of concepts (resp.~roles) of the form~$\pf D$ (resp.~$\pf R$):

\begin{definition}[Semantics]
A \df{bi-ordered interpretation} $\OO=\tuple{\Delta^{\OO},\cdot^{\OO},<^{\OO},\ll^{\OO}}$ interprets the classical constructors in the usual way. The typicality-based concepts are interpreted as $(\pf D)^{\OO}\defined\min_{<^{\OO}} D^{\OO}$, and~roles~as~$(\pf R)^{\OO}\defined\min_{\ll^{\OO}} R^{\OO}$.
\end{definition}
Hence, to be a typical element of a concept (resp.~role) amounts to being one of the most preferred elements in the interpretation of that concept (resp.~role). It is easy to see that the typicality operators are both \textit{idempotent}.
\medskip

The definition of \textit{satisfaction} of a statement~$\alpha$ by a bi-ordered interpretation~$\OO$, denoted as~$\OO\sats\alpha$, carries over from the classical case. If $X$ is a set of statements, with $\OO\sats X$, we denote the fact~$\OO$ satisfies each statement in~$X$, in which case we say~$\OO$ is a \textit{model} of~$X$. We say~$\OO$ is a model of a knowledge base~$\K=(\T,\R,\A)$, denoted~$\OO\sats\K$, if~$\OO$ is a model of ~$\T$, $\R$, and~$\A$.

Given a knowledge base~$\K$ and a statement~$\alpha$, we say~$\K$ \textit{preferentially entails}~$\alpha$, denoted $\K\models\alpha$, if, for every bi-ordered interpretation $\OO$ such that $\OO\sats\K$, we have $\OO\sats\alpha$.


\section{A formula representation}

The connection method represents the facts as matrices. In its classical approach to first-order logic, those facts are the so-called formulae, expressions that are evaluated to \true\ or \false according to the semantics of their operators. We see formulae in detail in this section.

\subsection{Vocabulary}

The vocabulary to build formulae contains a set \(\mathsf{V}\) for variables; for all integer \(n \geq 1\), a set of \(n\)-ary functions \(\mathsf{F}_n\); and the previously defined sets \CN, \RN, and \IN.
Additionally, we represent the relation between terms and pairs of terms with \(<\) and \(\ll\), respectively, in the literals of our representation. They are the syntactic counterparts of the preference relations on concepts and roles we have in the semantics.

\begin{definition}[Term]
A \df{term} is recursively defined as follows:
\begin{enumerate}
\item A variable is a term;
\item An individual name is a term;
\item Given a \(n\)-ary function \(f\), and terms \(t_1, \cdots, t_n\), \(f(t_1, \cdots, t_n)\) is a term.

With \(\mathsf{T}\), we represent the set of terms of a language.
\end{enumerate}
\end{definition}

\begin{definition}[Literal]
A \df{literal} is defined as follows:
\begin{equation*}
L ::= \CN(\mathsf{T}) \mid \RN(\mathsf{T}, \mathsf{T}) \mid \mathsf{T} < \mathsf{T} \mid (\mathsf{T}, \mathsf{T}) \ll (\mathsf{T}, \mathsf{T}) \mid \nao{L}
\end{equation*}
\end{definition}

\begin{definition}[Formula]
A \df{formula} is built upon the constructors \(\land\), \(\lor\), \(\neg\), \(\exists\), and \(\forall\) as follows:
\begin{equation*}
F ::= L \mid F \land F \mid F \lor F \mid \neg F \mid \exists \mathsf{V} F \mid \forall \mathsf{V} F
\end{equation*}
\end{definition}

\begin{definition}[Semantics]
An \df{interpretation} \(\II\) is a pair \(\langle \Delta^\II, \cdot^\II \rangle\), where \(\Delta^\II\) is the domain of interpretation, and \(\cdot^\II\) is the interpretation function that interprets formulae elements as: 
\begin{itemize}
\item \(a^\II \in \Delta^\II\), for each individual name \(a \in \mathsf{I}\);
\item \(A^\II : \Delta^\II \longrightarrow \{\true, \false\}\), for each concept name \(A \in \mathsf{C}\);
\item \(r^\II : \Delta^\II \times \Delta^\II \longrightarrow\{\true, \false\}\), for each role name \(r \in \mathsf{R}\);
\item \(<^\II : \Delta^\II \times \Delta^\II \longrightarrow\{\true, \false\}\);
\item \(\ll^\II : (\Delta^\II \times \Delta^\II) \times (\Delta^\II \times \Delta^\II) \longrightarrow\{\true, \false\}\);
\item \(f^\II : (\Delta^\II)^n \longrightarrow \Delta^\II\), for each \(n \geq 1\) and function \(f \in \mathsf{F}_n\).
\end{itemize}
\end{definition}

\begin{definition}[Satisfaction]
Given a formula \(F\) and an interpretation~\(\II\), we say that \(\II\) \df{satisfies} \(F\), denoted \(\II \sats F\), if:
\begin{itemize}
\item if \(F = \top\), then \(\top^\II = \true\);
\item if \(F = \bot\), then \(\bot^\II = \false\);
\item if \(F = A(t)\), then \(\II \sats F\) if \(A^\II(t^\II) = \true\);
\item if \(F = r(t, u)\), then \(\II \sats F\) if \(r^\II(t^\II, u^\II) = \true\);
\item if \(F = t < u\), then \(\II \sats F\) if \(<^\II(t^\II, u^\II) = \true\);
\item if \(F = (t, u) \ll (v, k)\), then \(\II \sats F\) if \(\ll^\II(t^\II, u^\II, v^\II, k^\II) = \true\);
\item if \(F = \nao F_1\), then \(\II \sats F\) if \(F_1^\II = \false\);
\item if \(F = F_1 \lor F_2\), then \(\II \sats F\) if \(\II \sats F_1\), or \(\II \sats F_2\), or both;
\item if \(F = F_1 \land F_2\), then \(\II \sats F\) if \(\II \sats F_1\) and \(\II \sats F_2\);
\item if \(F = \forall x F_1\), then \(\II \sats F\) if \(\II' \sats F_1\), for every extension \(\II'\) that coincides with \(\II\) but maps \(x\) to different objects \(x^\II\) of \(\Delta^\II\);
\item if \(F = \exists x F_1\), then \(\II \sats F\) if \(\II' \sats F_1\), for some extension \(\II'\) that coincides with \(\II\) but maps \(x\) to different objects \(x^\II\) of \(\Delta^\II\).
\end{itemize} 
\end{definition}

The definitions above recall the well-known classical first-order logic. In our case, the formulae are simpler than the full first-order once there is no functions (besides Skolem functions) or predicates with more than two terms.
\section{Mapping from \alchb\ to matrices}

Description Logics and FOL have a close relationship. In fact, families of DLs are decidable fragments of FOL \cite{basic.dl}. Baader et. al present a translation function from \alc\ knowledge bases to FOL formulae. We extend this translation to \alchb\ knowledge base, defining a translation function \(\pi\) such that maps roles, concepts and knowledge bases to FOL formulae. The idea is to denote the notion of typical elements into FOL predicates, adding new axioms to ensure those predicates represent strict partial orders.

\begin{definition}[Translation between roles and formulae]
Given the terms \(t, u\), the \df{translation function $\pi_{t, u}$} maps \alchb-roles to formulae with variables $x$ and $y$ as follows:
\begin{align*}
\pi_{t,u} (r) &\defined r(t,u), \\
\pi_{t,u} (\nao r) &\defined \nao{r(t,u)}, \\
\pi_{t,u} (\pf r) &\defined r(t,u) \land \forall x \forall y \left( \nao((x,y) \ll (t,u)) \lor \nao{r(x,y)}  \right), \\
\pi_{t,u} (\nao{\pf r}) &\defined \nao{r(t,u)} \lor \exists x \exists y \left( (x,y) \ll (t,u) \land \pi_{x,y} (\pf r)  \right). \\
\end{align*}
\end{definition}

\begin{definition}[Translation between concepts and formulae]
Given a term \(t\), the \df{translation function $\pi_{t}$} maps \alchb-concepts to formulae with a~variable~\(y\) as follows:
\begin{align*}
\pi_{t} (A) &\defined A(t) \\
\pi_{t} (D \e E) &\defined \pi_t (D) \land \pi_t (E) \\
\pi_{t} (D \ou E) &\defined \pi_t (D) \lor \pi_t (E) \\
\pi_{t} (\exists R . D) &\defined \exists y \left[\pi_{t, y} (R) \land \pi_y (D)\right] \\
\pi_{t} (\forall R . D) &\defined \forall y \left[ \pi_{t, y} (\nao{R}) \lor \pi_y (D)\right] \\
\pi_{t} (\pf D) &\defined \pi_t (D) \land \forall y \left[\nao{(y < t)} \lor \pi_y (\nao{D})\right] \\
\pi_{t} (\nao{\pf D}) &\defined \pi_t (\nao{D}) \lor \exists y \left[(y < t) \land \pi_y (\pf D)\right]
\end{align*}
\end{definition}

\begin{definition}[Translation between knowledge bases and formulae] \label{def:pi}
The \df{translation function $\pi$} maps a knowledge base $\K$ to formulae as follows:
\begin{align}
&\pi(\K) \defined 
\bigwedge_{a : D \in \K} \pi_{a}(D)
\land \bigwedge_{(a, b) : R \in \K} \pi_{a, b}(R)
\bigwedge_{D \sub E \in \K} \forall x \left[ \pi_{x}(\nao{D} \ou E)\right] \\
&\bigwedge_{R \sub S \in \K} \forall x \forall y \left[ \pi_{x,y}(\nao{R}) \lor \pi_{x,y}(E))\right] \\
&\land \forall x \forall y \forall z \left[\nao{(x < y \land y < z)} \lor x < z\right] \\
&\land \forall x \left[\nao{(x < x)}\right]\\
&\land \forall x \forall y \left[\nao{(x < y)} \lor \nao{(y < x)}\right] \\
&\land \forall x \forall y \forall z \forall k \forall m \forall n \left[\nao{((x, y) \ll (z,k)
\land (z, k) \ll (m, n))} \lor (x, y) \ll (m, n)\right] \\
&\land \forall x \forall y \left[\nao{((x, y) \ll (x, y))}\right] \\
&\land \forall x \forall y \forall z \forall k \left[\nao{((x, y) \ll (z, k))} \lor \nao{((z, k) \ll (x, y))}\right] 
\end{align}
\end{definition}

The previous definition maps each axiom to an atomic subformula (Eq. 1 and Eq. 2) from the conjunction defined above. The other conjunction's subformulae are not related to the knowledge base; they consist of transitivity (Eq. 3 and Eq. 6), irreflexivity (Eq. 4 and Eq. 7) and asymmetry (Eq. 5 and Eq. 8) axioms for \(<\) and \(\ll\), to ensure that such relations are strict partial orders.
Hereafter, we omit those axioms unless we use them in proof.

\begin{corollary} \label{corollary:II}
Given a knowledge base \(\K\) and a bi-ordered interpretation \(\OO\), if \(\OO \sats \K\), then there exists an interpretation \(\II\) such that \(\II \sats \pi(\K)\)
\end{corollary}

\begin{proof}
    This proof is a long one and, thus, is split into three parts. In the first part, we build an interpretation based on the bi-ordered interpretation \(\OO\). The second part contains the proofs for concept and role assertions. The last one is related to TBox axioms as such concept and role subsumption.
    
    \paragraph{Definition of an interpretation}
    
    To show this corollary, we start by building an interpretation \(\II\) such that:
    \begin{enumerate}
    \item \(\Delta^\II\) is the domain \(\Delta^\OO\);
    
    \item \(a^\II = a^\OO\), for all individual names \(a\);
    
    \item for each concept name \(A \in \CN\) and each object \(o \in \Delta^\OO\), \[A^\II(o) = \begin{cases}
    \true\text{, if }o \in A^\OO \\
    \false\text{, otherwise}
    \end{cases};\]
    
    \item for each role name \(r \in \RN\) and each pair of objects \((o_1, o_2) \in \Delta^\OO \times \Delta^\OO\), \[r^\II(o_1, o_2) = \begin{cases}
    \true\text{, if }(o_1, o_2) \in r^\OO \\
    \false\text{, otherwise}
    \end{cases};\]
    
    \item for each pair \((o_1, o_2) \in \Delta^\II \times \Delta^\II\)
    \[<^\II(o_1, o_2) = \begin{cases}
    \true\text{, if }(o_1, o_2) \in (<^\OO)^+ \\
    \false\text{, otherwise}
    \end{cases};~\text{and}\]
    
    \item for each pair of pairs of objects \(((o_1, o_2), (o_3, o_4)) \in (\Delta^\II \times \Delta^\II) \times (\Delta^\II \times \Delta^\II)\)
    \[\ll^\II((o_1, o_2), (o_3, o_4)) = \begin{cases}
    \true\text{, if }((o_1, o_2), (o_3, o_4)) \in (\ll^\OO)^+ \\
    \false\text{, otherwise}
    \end{cases}\]
    \end{enumerate}
    
    This interpretation is suitable once there is no \(A^\II(o)\) both \(\true\) and \(\false\), otherwise \(\OO\) could not satisfy \(\K\). Furthermore, \(\II\) satisfies the transitivity, irreflexivity, and asymmetry axioms for both \(<^\II\) and \(\ll^\II\) since \(<^\OO\) and \(\ll^\OO\) are strict partial orders, by definition.
    To illustrate those properties, assume that~\(\II\) does not satisfy the irreflexivity axiom for \(<\). Then, \(\II \not\sats \forall x . \nao{(x < x)}\), which means there exists an object \(o \in \Delta^\II\) such that \(<^\II(o, o) = \true\). However, by definition of \(\II\), \(<^\II(o, o)\) must be \false, as there is no \((o, o) \in <^\OO\) because \(<^\OO\) is a strict partial order. The other properties are obtained in a similar manner and are left to the reader.
    
    \paragraph{Proof for role assertions}
    
    We here prove satisfaction for role assertions by induction on the structure of formulae, such that
    \[[\pi_{a, b}(R)]^\II = \begin{cases}
    \true\text{, if }(a^\OO, b^\OO) \in R^\OO \\
    \false\text{, otherwise.}
    \end{cases}.\]
    
    The assertions which are related to role names or their negation constitute the induction base cases as follows:
    \begin{itemize}
    \item \(R = r\). Then, \((a^\OO, b^\OO) \in r^\OO\) and, by definition of \(\II\) above, \(r^\II(a^\II, b^\II) = \true\); and
    
    \item \(R = \nao{r}\). Then, \((a^\OO, b^\OO) \not\in r^\OO\) and, by definition of \(\II\) above, \(r^\II(a^\II, b^\II) = \false\).
    \end{itemize}
    
    Induction steps:
    \begin{itemize}
    \item \(R = \nao{S}\). Then, \((a^\OO, b^\OO) \not\in S^\OO\). By the IH, \([\pi_{a, b}(S)]^\II = \false\). Therefore, \([\pi_a(\nao{S})]^\II = \true\);
    
    \item \(R = \pf{S}\). Then, \((a^\OO, b^\OO) \in S^\OO\) and, for every pair \(o_1, o_2 \in \Delta^\OO \times \Delta^\OO\), either \(((o_1, o_2), (a^\OO, b^\OO)) \not\in (\ll^\OO)^+\), or \((o_1, o_2) \in (\nao{S})^\OO\). By the IH, the definition of \(\II\), and the semantics of formula, we have that \([\pi_{a, b}(S)]^\II = \true\), and \linebreak\([\forall x \forall y . \nao{((x, y) \ll (a, b)) \lor \pi_{x, y}(\nao{S})}]^\II = \true\) as well. Therefore, \linebreak \([\pi_{a, b}(S) \land \forall x \forall y . \nao{((x, y) \ll (a, b)) \lor \pi_{x, y}(\nao{S})}]^\II = \true\) even as \([\pi_{a, b}(\pf S)]^\II = \true\);
    
    \item \(R = \nao{\pf{S}}\). Then, \((a^\OO, b^\OO) \not\in S^\OO\) or, for some pair \(o_1, o_2 \in \Delta^\OO \times \Delta^\OO\), \(((o_1, o_2), (a^\OO, b^\OO)) \in (\ll^\OO)^+\) and \((o_1, o_2) \in S^\OO\).
    By the IH, the definition of \(\II\), and the semantics of formula, we have that \([\pi_{a, b}(\nao{S})]^\II = \true\), or \linebreak\([\exists x \exists y . ((x, y) \ll (a, b)) \land \pi_{x, y}(S)]^\II = \true\), or both. Therefore, \linebreak \([\pi_{a, b}(\nao{S}) \lor \exists x \exists y . ((x, y) \ll (a, b)) \land \pi_{x, y}(S)]^\II = \true\), meaning that \([\pi_{a, b}(\nao{\pf S})]^\II = \true\), too.
    
    \end{itemize}
    
    \paragraph{The proof for concept assertions}
    
    We here prove the satisfaction for concept assertions, such that
    
    \[[\pi_a(D)]^\II = \begin{cases}
    \true\text{, if }a^\OO \in D^\OO \\
    \false\text{, otherwise}
    \end{cases}\]
    
    The assertion related to concept names is the induction base cases. Then, \(a^\OO \in A^\OO\) and, by definition of \(\II\) above, \(A^\II(a^\II) = \true\).
    
    Induction steps:
    \begin{itemize}
    \item \(D = \nao{E}\). Then, \(a^\OO \in \Delta^\OO \setminus E^\OO\) which means that \(a^\OO \not\in E^\OO\). By the IH, \([\pi_a(E)]^\II = \false\). Therefore, \([\pi_a(\nao{E})]^\II = \true\);
    
    \item \(D = {E_1 \ou E_2}\). Then, \(a^\OO \in E_1^\OO \cup E_2^\OO\) which means that either \(a^\OO \in E_1^\OO\), \(a^\OO \in E_2^\OO\), or both. By the IH, either \([\pi_a(E_1)]^\II = \true\), \([\pi_a(E_2)]^\II = \true\), or both. Therefore, \([\pi_a(E_1 \ou E_2)]^\II = \true\);
    
    \item \(D = {E_1 \e E_2}\). Then, \(a^\OO \in E_1^\OO \cap E_2^\OO\) which means that \(a^\OO \in E_1^\OO\) and \(a^\OO \in E_2^\OO\). By the IH, \([\pi_a(E_1)]^\II = \true\) and \([\pi_a(E_2)]^\II = \true\). Therefore, \([\pi_a(E_1 \e E_2)]^\II = \true\);
    
    \item \(D = {\exists R . E}\). Then, there exists an object \(x\) such that \((a^\OO, x) \in R^\OO\) and  \(x \in E^\OO\). By the IH, \([\exists y .\pi_{a, y}(R) \land \pi_y(E)]^\II = \true\). Therefore, \([\pi_a(\exists R . E)]^\II = \true\);
    
    \item \(D = {\forall R . E}\). Then, for all objects \(x\), \((a^\OO, x) \not\in R^\OO\) or  \(x \in E^\OO\). By the IH, \([\forall y .\pi_{a, y}(\nao{R}) \lor \pi_y(E)]^\II = \true\). Therefore, \([\pi_a(\forall R . E)]^\II = \true\);
    
    \item \(D = \pf{E}\). Then, \(a^\OO \in E^\OO\) and, for all object \(o \in \Delta^\OO\), either \((o, a^\OO) \not\in (<^\OO)^+\), or \(o \in (\nao{E})^\OO\). By the IH, the definition of \(\II\), and the semantics of formula, we have that \([\pi_{a}(E)]^\II = \true\), and \linebreak[2]\([\forall x . \nao{(x < a) \lor \pi_{x}(\nao{E})}]^\II = \true\) as well. Therefore, \linebreak[2]\([\pi_{a}(E) \land \forall x . \nao{(x < a) \lor \pi_{x}(\nao{E})}]^\II = \true\), also \([\pi_{a}(\pf E)]^\II = \true\) too;
    
    \item \(D = \nao{\pf{E}}\). Then, \(a^\OO \not\in E^\OO\) or, for some object \(o \in \Delta^\OO\), \linebreak\((o, a^\OO) \in (<^\OO)^+\) and \(o \in E^\OO\).
    By the IH, the definition of \(\II\), and the semantics of formula, we have that \([\pi_{a}(\nao{E})]^\II = \true\), \linebreak[2]\([\exists x. x < a \land \pi_{x}(E)]^\II = \true\), or both. Therefore, \linebreak[2]\([\pi_{a}(\nao{E}) \lor \exists x . x < a \land \pi_{x}(E)]^\II = \true\), meaning that \([\pi_{a}(\nao{\pf E})]^\II = \true\) too.
    
    \end{itemize}
    
    \paragraph{The proof for role subsumption}
    We prove that, given an axiom \(R \sub S \in \K\),
    \[
    [\forall x \forall y [\pi_{x, y}(\nao{R}) \lor \pi_{x, y}(S)]]^\II = \begin{cases}
    \true,~\text{if}~R^\OO \subseteq S^\OO \\
    \false,~\text{otherwise}
    \end{cases}
    \]
    by contrapositive.
    Assume that there exists a pair of objects \((o_1, o_2) \in \Delta^\OO \times \Delta^\OO\), such that \((o_1, o_2) \in R^\OO\), but \((o_1, o_2) \not\in S^\OO\). Let them be two new individual names \(i_1\) and \(i_2\). We show above that the translation function holds for individual names. Then, \([\pi_{i_1, i_2}(\nao{S}) \lor \pi_{i_1, i_2}(S)]^\II = \false\). By the definition of the satisfaction of \(\forall\), a formula is interpreted as \true\ only if every object in the domain is also interpreted as \true. Therefore, \([\forall x \forall y. \pi_{x, y}(\nao{R}) \lor \pi_{x, y}(S)]^\II = \false\).
    
    \paragraph{The proof for concept subsumption}
    Here, we prove that, given an axiom \(D \sub E \in \K\),
    \[
    [\forall x. \pi_{x}(\nao{D} \ou E)]^\II = \begin{cases}
    \true,~\text{if}~D^\OO \subseteq E^\OO \\
    \false,~\text{otherwise}
    \end{cases}
    \]
    by contrapositive.
    Assume that there exists an object \(o\), such that \(o \in D^\OO\), but \(o \not\in E^\OO\). Let it be a new individual name \(i\). We prove above that the translation function for concept expression holds for individual names. Then, \([\pi_{i}(\nao{D} \ou E)]^\II = \false\). By the definition of the satisfaction of \(\forall\), the formula is interpreted as \true\ only if, for every object in the domain, it is also interpreted as \true. Therefore, \([\forall x. \pi_{x}(\nao{D} \ou E)]^\II = \false\).
    
\qed\end{proof}


\begin{corollary} \label{corollary:OO}
Given a knowledge base \(\K\) and an interpretation \(\II\), if \(\II \sats \pi(\K)\), then there exists a bi-ordered interpretation \(\OO\) such that \(\OO \sats \K\).
\end{corollary}

\begin{proof}
    As the previous corollary, this proof is quite long and contains three main parts. In the first part we build a bi-ordered interpretation based on the interpretation \(\II\). The second part contains the proofs for concept and role assertions. Lastly, the proof is related to TBox axioms as such concept and role subsumption.
    
    We prove this corollary by creating a bi-ordered interpretation such that it satisfies the knowledge base \(\K\). So, given the interpretation \(\II\), we build a bi-ordered interpretation, where:
    \begin{itemize}
    \item \(\Delta^\OO = \Delta^\II\);
    
    \item for each individual name \(a\), \(a^\OO = a^\II\);
    
    \item for each concept name \(A\) and object \(o \in \Delta^\II\), \(o \in A^\OO\) if \(A^\II(o) = \mathsf{true}\);
    
    \item for each role name \(r\) and objecs \(o_1, o_2 \in \Delta^\II\), \((o_1, o_2) \in r^\OO\) if \(r^\II(o_1, o_2) = \mathsf{true}\);
    
    \item for each pair of objects \(o_1, o_2 \in \Delta^\II\), \((o_1, o_2) \in <^\OO\) if \(<^\II(o_1, o_2) = \mathsf{true}\); and
    
    \item for each pair of pairs of objects \(o_1, o_2, o_3, o_4 \in \Delta^\II\), \(((o_1, o_2), (o_3, o_4)) \in \ll^\OO\) if \(\ll^\II(o_1, o_2, o_3, o_4) = \mathsf{true}\).
    \end{itemize}
    
    
    \paragraph{Proof for preferential relations}
    
    First, \(<^\OO\) and \(\ll^\OO\) must be strict partial orders, e.g., transitive, irreflexive and asymmetric. We prove for \(<^\OO\) by contrapositive, but the proof for \(\ll^\OO\) is analogous and left to the reader. We assume that \(<^\OO\) is not a strict partial order. Then, the following cases may happen:
    \begin{enumerate}
    \item there exists a pair \((o, o) \in <^\OO\);
    \item there exists \((o_1, o_2) \in <^\OO\) and \((o_2, o_3) \in <^\OO\), but \((o_1, o_3) \not\in <^\OO\); or
    \item there exists  \((o_1, o_2) \in <^\OO\) and \((o_2, o_1) \in <^\OO\).
    \end{enumerate}
    If case 1 is true, then \(\II \sats  \exists x (x < x)\), by the Definition of \(\OO\). Hence, \(\II \not\sats  \forall x \nao{(x < x)}\).
    If case 2 is true, then \(\II \sats \exists x \exists y \exists z \left[x < y \land y < z \land \nao{(x < z)}\right]\), by the definition of \(\OO\). Hence, \(\II \not\sats \forall x \forall y \forall z \left[\nao{(x < y \land y < z)} \lor x < z\right]\).
    If case 3 is true, then \(\II \sats \exists x \exists y \left[(x < y) \land (y < x)\right]\), by the definition of \(\OO\).
    Hence, \(\II \not\sats \forall x \forall y \left[\nao{(x < y)} \lor \nao{(y < x)}\right]\).
    Therefore, \(<^\OO\) is a strict partial order.
    
    \paragraph{Proof for role assertions}
    
    We prove that \(\OO\) satisfies the role assertions in~\(\K\) by showing that if \([\pi_{a, b}(R)]^\II = \true\), then \((a^\OO, b^\OO) \in R^\OO\) by induction over role expressions.
    There are three basis cases:
    \begin{itemize}
    
    \item Case \(R\) is a role name \(r\) or its negation. Then, by definition of \(\II\), if \([r(a, b)]^\II = \true\), then \((a^\OO, b^\OO) \in r^\OO\), and if \([\nao{r}(a, b)]^\II = \true\), then \([r(a, b)]^\II = \false\), therefore, \((a^\OO, b^\OO) \not\in r^\OO\);
    \item Case \(R = \pf r\). Then, \(\pi_{a, b}(\pf r)\) is \(\left[r(a,b) \land \forall x \forall y \left[\nao{((x,y) \ll (a,b))} \lor \nao{r}(x, y)\right]\right]^\II = \true\). Hence, \([r(a, b)]^\II = \true\), and for all pairs of objects in \(\Delta^\II\), either \([(x,y) \ll (a,b)]^\II = \true\) or \([\nao{r}(x, y)]^\II = \true\). So, by the definition of \(\OO\), \((a^\OO, b^\OO) \in r^\OO\) and, either \(((o_1,o_2), (a^\OO,b^\OO)) \not\in (\ll^\OO)^+\) or \((o_1, o_2) \not\in r^\OO\), for all \(o_1, o_2 \in \Delta^\OO\). Therefore, \((a^\OO, b^\OO) \in (\pf r)^\OO\).
    \end{itemize}
    
    The induction step is for \(R = \nao{\pf} r\). Then, \(\pi_{a, b}(\nao{\pf r})\) is \(\left[\nao{r}(a,b) \lor \exists x \exists y \left[(x,y) \ll (a,b) \land \pi_{x,y}(\pf r)\right]\right]^\II = \true\). Hence, by the semantics of intepretation, either \([\nao{r}(a, b)]^\II = \true\) or, for some pair of objects in \(\Delta^\II\), \([((x,y) \ll (a,b))]^\II = \true\) and \([\pi_{x, y}(\pf r)]^\II = \true\). So, by the definition of \(\OO\) and the IH, either \((a^\OO, b^\OO) \not\in r^\OO\), or \(((o_1,o_2), (a^\OO,b^\OO)) \in (\ll^\OO)^+\) and \((o_1, o_2) \in (\pf r)^\OO\), for some \(o_1, o_2 \in \Delta^\OO\). Therefore, \((a^\OO, b^\OO) \in (\nao{\pf} r)^\OO\).
    
    \paragraph{Proof for concept assertions}
    
    Here, we prove that if \((\pi_a{D})^\II = \true\), then \(a^\OO \in D^\OO\) by induction over concept expressions in \alchb.
    
    The induction basis case is the one where \(D\) is a concept name. In this case, \(\pi_a(A) = A(a)\), meaning that \(A^\II(a^\II) = \true\). By the definition of \(\OO\), \(a^\OO \in A^\OO\).
    
    Induction steps:
    \begin{itemize}
    \item Case \(D = \nao{E}\). Then, \((\pi_a(\nao{E}))^\II = \true\). So, \([\pi_a(E)]^\II = \false\), and by the IH, \(a^\OO \not\in E^\OO\). Therefore, \(a^\OO\in (\nao{E})^\OO\);
    
    \item Case \(D = E_1 \ou E_2\). Then, either \([\pi_{a}(E_1)]^\II = \true\), or \([\pi_{a}(E_2)]^\II = \true\), or both. By the IH, \(a^\OO \in E_1^\OO \cup E_2^\OO\). Therefore, \(a^\OO \in (E_1 \ou E_2)^\OO\);
    
    \item Case \(D = E_1 \e E_2\). Then, \([\pi_{a}(E_1)]^\II = \true\) and \([\pi_{a}(E_2)]^\II = \true\). By the IH, \(a^\OO \in E_1^\OO \cap E_2^\OO\). Therefore, \(a^\OO \in (E_1 \e E_2)^\OO\);
    
    \item Case \(D = \exists R . E\). Then, \([\exists x \left[ \pi_{a, x}(R) \land \pi_{a}(E)\right]]^\II = \true\). By the IH and the semantics, there exists an object \(o \in \Delta^\OO\) such that \((a^\OO, o) \in R^\OO\) and \(o \in E^\OO\). Therefore, \(a^\OO \in (\exists R.E)^\OO\);
    
    \item Case \(D = \forall R . E\). Then, \([\forall x \left[\pi_{a, x}(\nao{R}) \lor \pi_{a}(E)\right]]^\II = \true\). By the IH and the semantics, for all objects \(o \in \Delta^\OO\), either \((a^\OO, o) \not\in R^\OO\), or \(o \in E^\OO\). Therefore, \(a^\OO \in (\forall R.E)^\OO\);
    
    \item Case \(D = \pf E\). Then \([\pi_{a}(E) \land \forall x \left[\nao{(x < a)} \lor \pi_{a}(\nao{E})\right]]^\II = \true\). By the IH and the semantics, for all objects \(o \in \Delta^\OO\), either \((o, a^\OO) \not\in (<)^\OO\), or \(o \in (\nao{E})^\OO\). Therefore, \(a^\OO \in (\pf E)^\OO\); and
    
    \item Case \(D = \nao{\pf} E\). Then \([\pi_{a}(\nao{E}) \lor \forall x \left[(x < a) \land \pi_{a}(\pf E)\right]]^\II = \true\). By the IH and the semantics, for some object \(o \in \Delta^\OO\), \((o, a^\OO) \in (<)^\OO\), and \(o \in (\pf{E})^\OO\). Therefore, \(a^\OO \in (\nao{\pf} E)^\OO\).
    
    \end{itemize}
    
    \paragraph{Proof for role subsumptions}
    
    We prove \(\OO\) satisfies role subsumptions by contrapositive. Assume that \([\forall x \forall y. \pi_{x, y}(\nao{R}) \lor \pi_{x, y}(S)]^\II = \false\). Then, there exists a pair of objects \(o_1, o_2\) such that the formula is interpreted as \false. Let them be two new individual names \(i_1\), and \(i_2\) such that \(i_1^\II = o_1\) and \(i_2^\II = o_2\). Hence, \([\pi_{i_1, i_2}(\nao{R}) \lor \pi_{i_1, i_2}(S)]^\II = \false\), and, by the proof for role assertions and satisfaction of \(\nao{}\), \((o_1, o_2) \in R^\OO\), but \((o_1, o_2) \not\in S^\OO\). Therefore, \(R^\OO \not\subseteq S^\OO\).
    
    \paragraph{Proof for concept subsumptions}
    
    We prove \(\OO\) satisfies concept subsumptions by contrapositive. Assume that \([\forall x. \pi_{x}(\nao{D} \ou E)]^\II = \false\). Then, there exists an object \(o\) such that the formula is interpreted as \false. Let it be a new individual name \(i\), such that \(i^\II = o\). Hence, \([\pi_{i}(\nao{D} \ou E)]^\II = \false\), and, by the proof for concept assertions and satisfaction of \(\nao{}\), \(o \in D^\OO\), but \(o \not\in E^\OO\). Therefore, \(D^\OO \not\subseteq E^\OO\).
    
\qed\end{proof}


\section{From formulae to matrices}

In the previous section, we show the translation from any knowledge base to formulae. 
Those formulae are fragments of the so-called first-order logic (FOL) used in various applications and automated theorem-proving algorithms.
The connection method is one prominent example of such algorithms. The intuition for the connection method is its representation of formulae as matrices, extending the proof search connecting clauses of the matrix in order to check whether some formula is valid.
The connection method (CM)~\cite{bibel:cm} consists in a validity procedure (in opposition to refutation procedures such as tableaux and resolution), i.e., it tries to prove whether a formula (query) is valid directly. 
If $X=\{\varphi_{1},\ldots,\varphi_{n}\}$ is a (finite) set of first-order formulae (also referred to as a knowledge base, $\KB$), then in order to check whether $X\models\psi$, for some first-order formula~$\psi$, the validity of the formula $(\varphi_{1}\land\ldots\land\varphi_{n})\to\psi$ ($X\to\psi$), i.e., of $\nao{X}\lor\psi$, must be proven.
Hence, CM requires the conversion of formulae into the disjunctive normal form (DNF).
The consequences of negating the $\KB$ are the following: (\emph{i}) subsumption axioms (e.g., $E \sub D$) are converted into $E\land \nao{D}$; (\emph{ii}) variables are existentially quantified, instead of universally; (\emph{iii}) thus, FOL Skolemization applies over universally quantified variables, and (\emph{iv}) the consequent (or query) $\alpha$ is not negated.
Of course, when moving to the DL case, the crux of the matter is precisely how to express the negation of the knowledge base, along with doing away with variables and Skolem functions.

\alctcm\ \cite{fred:calculus} was the first CM proposed for DLs. Among its main features is the fact it requires neither variables in the initial representation of the axioms, nor Skolem functions. Moreover, it includes a blocking solution to ensure termination, as commonly done in the field, and the definition of $\theta$-substitution as a suitable replacement for variable unification.
Furthermore, a connection calculus for a defeasible extension of DLs has been proposed earlier~\cite{alcbtcm} for a less expressive language \alcb. It handles the preference relation as an auxiliary structure and relies on a normal form, the Typicality Normal Form.

\begin{lemma}
Given a knowledge base \(\K\) and an axiom \(\alpha\), \(\K \models \alpha\) iff \(\models \nao{\pi{(\K)}} \lor \pi(\alpha)\).
\end{lemma}

\begin{proof}
We prove this lemma by contrapositive on both sides.

\paragraph{Only-if part:} Assume that \(\not\models \nao{\pi{(\K)}} \lor \pi(\alpha)\). Then, there exists an interpretation \(\II\) such that \([\nao{\pi{(\K)}} \lor \pi(\alpha)]^\II = \false\), meaning that \([\pi{(\K)} \land \nao{\pi(\alpha)}]^\II = \true\). It is already proven, by the Corollary \ref{corollary:OO}, that we can build a bi-ordered interpretation \(\OO\) such that \(\OO \sats \K\) but \(\OO \not\sats \alpha\). Therefore, \(\K \not\models \alpha\).

\paragraph{If part:} Assume that there exists a bi-ordered interpretation \(\OO\) such that \(\OO \sats \K\) but \(\OO \not\sats \alpha\). It is already proven, by the Corollary \ref{corollary:II}, that we can build an interpretation \(\II\), such that \([\nao{\pi{(\K)}} \lor \pi(\alpha)]^\II = \false\). Therefore, \(\not\models \nao{\pi{(\K)}} \lor \pi(\alpha)\)
\qed\end{proof}

As a fragment of FOL, some useful results such as the so-called negation normal form (NNF), Prenex normal form (PNF), Skolem normal form (SNF) and their equivalence hold here as well.
However, the Skolemization preserves the consistency of formulae but not the validity of them. 
Hence, as the connection method is a validity procedure, we use the dual version of Skolemization, the \df{Herbrandization}.
In this process, the formulae contain only \(\exists\), replacing the \(\forall\) for functions.

\begin{definition}[Prenex Normal Form (PNF)]
A formula \(F\) is in \df{Prenex normal form} if
\(F = Q_1 x_1 Q_2 x_2 \cdots Q_n x_n ~ \varphi\) 
where, for all \(1 \leq i \leq n\), \(Q_i \in \{\forall, \exists\}\) and \(\varphi\) is a quantifier-free formula.
\end{definition}

The FOL fragment we use here contains variables only \df{guarded} by quantifiers, e.g., no quantifier-free variables exist. As such, dealing with free variables is not the case, and the translation from any formula to its PNF is straightforward.

\begin{definition}[Herbrand Normal Form (HNF)]
A formula is in \df{Herbrand Normal Form}\footnote{The HNF is a dual normal form of Skolem Normal Form, as Herbrandization is the dual process to remove quantifiers as Skolemization process as well.} if it is in PNF, and for all \(1 \leq i \leq n\), \(Q_i = \exists\).
\end{definition}

In other words, there are no universal quantifiers on HNF. It is well-known that every formula in FOL can be converted to a formula in HNF due to a process called \df{Herbrandization}. It ensures that both formulae are equivalent: one is valid iff the other is valid as well.
Hereafter we shall assume that formulae are quantifier-free, where the remaining variables are existentially quantified, and Skolem functions represent previous universally quantified variables.

\begin{definition}[Disjunctive Normal Form (DNF)]
A formula is in \df{Disjunctive normal form} if it is a disjunction of conjunctions of literals.
\end{definition}

\begin{corollary}
Given a formula \(F\) and an interpretation \(\II\), there exists a formula \(F'\) in DNF, such that \(\II \sats F\) iff \(\II \sats F'\).
\end{corollary}

The proof of this corollary comes from the facts that first-order operators are distributive.

\begin{definition}[Validity]
Given a formula \(F\), we say it is \df{valid}, denoted \(\models F\), if \(\II \sats F\), for every interpretation \(\II\).
\end{definition}

\begin{definition}[Matrix]
Given a formula \(F = \exists x_1 \cdots \exists x_o  D_1 \lor \cdots \lor D_n\), where each \(D_i = L_{i,1} \land \cdots \land L_{i,m}\) and \(L_{i,j}\) is a literal, a \df{matrix} is the set of clauses \(\{C_1, \cdots, C_n\}\) where each clause \(C_i = \{L_{i,1}, \cdots , L_{i,m}\}\) is the set of literals of each \(D_i\).
\end{definition}

In other words, a matrix is the clausal form of a DNF formula without its quantifiers. So, given an interpretation \(\II\) and a matrix \(M = \{C_1, \cdots, C_n\}\), \(\II \sats M\) iff \(\II \sats F\), where \(F = \exists x_1 \cdots \exists x_o D_1 \lor \cdots \lor D_n\), and each \(D_i = L_{i,1} \land \cdots \land L_{i,m}\).

\subsection{A shortcut from \alchb\ knowledge bases to matrices}

In the previous sections, we define and prove that a mapping function exists from \alchb\ to a fragment of first-order logic.
Moreover, from the formulae in first-order logic, we show a matrix which is the clausal form of the formula in DNF (and after Herbrandization) in this section.
In detail, to map a knowledge base to a matrix, the following steps must occur:
\begin{enumerate}
\item to map the knowledge base to a formula in first-order logic;
\item to apply the negation of the formula in order to prove validity;
\item to transform the formula into Prenex normal form;
\item to apply the Herbrandization;
\item to transform the formula into DNF; and
\item to transform the conjunctions of the formula in DNF to its clausal form (the matrix).
\end{enumerate}

In this subsection, we provide a direct mapping from \alchb\ to matrices to obtain a more straightforward translation for knowledge bases, applying those six steps above in one translation.

At this point, we have a translation from \alchb\ to a formula (step 1). Now, we define a useful translation from formulae to matrices, encompassing steps 3, 4, 5, and 6. This translation is based on the matrix representation for formulae presented in Otten, 2012. \cite{jens:modal}

\begin{definition}[Clausal union] \label{def:clausal.union}
    Let $M_1$ and $M_2$ be matrices. We define the \df{clausal union} of $M_1$ and $M_2$ as $M_1 \clup M_2 \defined \{C_1 \cup C_2 \mid (C_1,C_2) \in M_1 \times M_2\}$.
\end{definition}

\begin{definition}[Application of a function symbol]
Given a function \(f\), an individual name \(c\), and a string of variables \(S = x_1 \cdots x_n\), we define the \df{application of} the function \(f\) with arguments \(S\) as:
\[
\mathsf{apply}(f, S) = \begin{cases}
c, \text{if } S = \varepsilon \\
f(x_1, \cdots, x_n), \text{otherwise}
\end{cases}
\]
\end{definition}

\begin{definition}[Matrix of a formula] \label{def:M}
Given a formula \(F\), a string of variables \(S\), a variable \(x\), and a function \(f\), the \df{matrix} of the formula \(F\), denoted \(M(F)\) is defined as follows:
\begin{itemize}
\item \(M(L, S) \defined \{\{L\}\}\);
\item \(M(F_1 \lor F_2, S) \defined M(F_1, S) \cup M(F_2, S)\);
\item \(M(F_1 \land F_2, S) \defined M(F_1, S) \clup M(F_2, S)\);
\item \(M(\exists x F_1, S) \defined M(F_1, S \cdot x)\);
\item \(M(\forall x F_1, S) \defined M(F_1, S)[x \mapsto \mathsf{apply}(f', S)]\).
\end{itemize}
\end{definition}

This translation from formulae to matrices already converts the whole formula to a formula in DNF and HNF equivalent one.

\begin{corollary}
Given a formula \(F\), \(M(F, \varepsilon)\) is the clausal form of \(F'\), such that \(F'\) is the formula \(F\) in DNF and HNF.
\end{corollary}

\begin{proof}
We prove this corollary by induction over the structure of the formula \(F\).
The induction basis case is when the formula is a literal. Therefore, \(M(F)\) is in DNF, HNF, and is equivalent to \(F\).

\paragraph{Induction steps:}
\begin{itemize}
\item When \(F = F_1 \lor F_2\). Then, by the IH, \(M(F_1, \varepsilon)\) is the formula \(F_1\) in DNF, HNF, and equivalent to \(F_1\), and \(M(F_2, \varepsilon)\) the same for \(F_2\). Therefore, the union of two DNF clausal form formulae is already a DNF clausal form for \(F_1 \lor F_2\);

\item When \(F = F_1 \land F_2\). Then, by the IH, \(M(F_1, \varepsilon)\) is the formula \(F_1\) in DNF, HNF, and equivalent to \(F_1\), and \(M(F_2, \varepsilon)\) the same for \(F_2\). Let \(F_1' = D_1 \lor \cdots \lor D_n\) and \(F_2' = E_1 \lor \cdots \lor E_m\) be the formulae of \(M(F_1, \varepsilon)\) and \(M(F_2, \varepsilon)\), respectively. To convert the conjunction of two formulae in DNF we apply the distributivity property for \(\lor\) and \(\land\) as follows:
\begin{align*}
F_1' \land F_2' = (D_1 \land E_1) \lor &\cdots \lor (D_1 \land E_m) \\
\lor (D_2 \land E_1) \lor &\cdots \lor (D_2 \land E_m) \\
&~~\vdots \\
\lor (D_n \land E_1) \lor &\cdots \lor (D_n \land E_m)
\end{align*}
which coincides with the cartesian product of their clauses. Therefore, \(M(F_1, \varepsilon)\) \(\clup\) \(M(F_2, \varepsilon)\) is in DNF, HNF and is equivalent to \(F_1 \land F_2\);

\item When \(F = \exists x F_1\). Then, by the IH, \(M(F_1, x)\) is the clausal form of the formula \(F_1\). As the matrix is the clausal form of a formula in HNF, it is easy to see that \(\exists x F'\) is equivalent to \(F'(x)\), as the variables are guarded by existential quantifiers, by the definition of HNF;

\item When \(F = \forall x F_1\). Then, by the IH, \(M(F_1, \varepsilon)\) is the clausal form of the formula \(F_1\). As the matrix is the clausal form of a formula in HNF, \(\forall x F'\) is equivalent to \(F'(\mathsf{apply(f, S)})\), where \(S\) is the set of variables before \(\forall\), as the Herbrandization process transform universally-restricted variables to functions such that their validity hold.
\end{itemize}
\end{proof}

\begin{definition}[Translation function for role expressions]
Given a role \(R\), two terms \(t, u\), two variables \(x, y\), and two functions \(f_i, g_i\), the \df{translation function $\varrho(R, t, u)$} maps role expressions to matrices as follows:
\begin{itemize}
\item \(\varrho(r, t, u) \defined \{\{r(t,u)\}\}\);
\item \(\varrho(\nao{r}, t, u) \defined \{\{\nao{r}(t, u)\}\}\);

\item \(\varrho(\nao{\pf r}, t, u) \defined \{\{\nao{r}(t,u)\}\} \cup \left[ \{\{(x, y) \ll (t, u)\}\} \clup \{\{r(x,y)\}\} \right]\);
\item \(\varrho(\pf r, t, u) \defined \{\{r(t,u)\}\} \clup \left[\{\{\nao{((f(t), g(u)) \ll (t, u))}\}\} \cup \varrho(\nao{\pf r}, f(t), g(u)) \right]\);

\item \(\varrho(\nao{\nao{R}}, t, u) \defined \varrho(R, t, u)\).

\end{itemize}
\end{definition}

\begin{definition}[Translation function for concept expressions]
Given a concept \(D\), a term \(t\), and a ordered set of terms \(S\), the \df{translation function $\varrho(D, t, S)$} maps concept expressions to matrices as follows:
\begin{enumerate}
\item \(\varrho(A, t, S) \defined \{\{A(t)\}\}\);
\item \(\varrho(\nao{A}, t, S) \defined \{\{\nao{A}(t)\}\}\);

\item \(\varrho(D \e E, t, S) \defined \varrho(D, t, S) \clup \varrho(E, t, S)\);
\item \(\varrho(\nao{(D \e E)}, t, S) \defined \varrho(\nao{D}, t, S) \cup \varrho(\nao{E}, t, S)\);

\item \(\varrho(D \ou E, t, S) \defined \varrho(D, t, S) \cup \varrho(E, t, S)\);
\item \(\varrho(\nao{(D \ou E)}, t, S) \defined \varrho(\nao{D}, t, S) \clup \varrho(\nao{E}, t, S)\);

\item \(\varrho(\exists R . D, t, S) \defined \varrho(R, t, x) \clup \varrho(D, x, S \cdot x)\);
\item \(\varrho(\nao{(\exists R . D)}, t, S) \defined \varrho(\nao{R}, t, \mathsf{apply}(f, S)) \cup \varrho(\nao{D}, \mathsf{apply}(f, S), S)\);

\item \(\varrho(\forall R . D, t, S) \defined \varrho(\nao{R}, t, \mathsf{apply}(f, S)) \cup \varrho(D, \mathsf{apply}(f, S), S)\);
\item \(\varrho(\nao{(\forall R . D, t, S)}) \defined \varrho(R, t, x) \clup \varrho(\nao{D}, x, S \cdot x)\);

\item \(\varrho(\pf D, t, S) \defined \varrho(D, t, S) \clup \left[\{\{\nao{(\mathsf{apply}(f, S) < t)}\}\} \cup \varrho(\nao{\pf D}, \mathsf{apply}(f, S), S) \right]\);
\item \(\varrho(\nao{\pf D}, t, S) \defined \varrho(\nao{D}, t, S) \cup \left[ \{\{x < t\}\} \clup \varrho(D, x, S \cdot x) \right]\);

\item \(\varrho(\nao{\nao{D}}, t, S) \defined \varrho(D, t, S)\).
\end{enumerate}
\end{definition}

\begin{definition}[Translation function for knowledge bases] \label{def:delta}
Given a knowledge base \(\K\) we define the \df{translation function $\delta$} from \(\K\) to a matrix as
\begin{align}
&\delta(\K) \defined \bigcup_{D \sub E \in \K} \varrho(D \e \nao{E}, x, x) 
\cup \bigcup_{a : D \in \K} \varrho(\nao{D}, a, \varepsilon) \\
&\cup \bigcup_{R \sub S \in \K} \left[\varrho(R, x, y) \clup \varrho(\nao{S}, x, y)\right] 
\cup \bigcup_{(a, b) : R} \varrho(\nao{R}, a, b) \\
&\cup \{\{x < y, y < z, \nao{(x < z)}\}\} \\
&\cup \{\{x < x\}\} \\
&\cup \{\{x < y, y < x\}\} \\
&\cup \{\{(x, y) \ll (z, k), (z, k) \ll (m, n), \nao{((x, y) \ll (m, n))}\}\} \\
&\cup \{\{(x, y) \ll (x, y)\}\} \\
&\cup \{\{(x, y) \ll (z, k), (z, k) \ll (x, y)\}\}
\end{align}
\end{definition}

\begin{definition}[Matrices equivalence]
Given two formulas \(F_1\) and \(F_2\), and their matrices \(M_1\) and \(M_2\), respectively. We say \(M_1\) and \(M_2\) are \df{equivalent} if, for every interpretation \(\II\), \(\II \sats F_1\) iff \(\II \sats F_2\).
\end{definition}

\begin{corollary}
Given a knowledge base \(\K\), \(M(\nao{\pi(\K)})\) and \(\delta(\K)\) are equivalent.
\end{corollary}

\begin{proof}
The proof of this corollary comes from the Definitions \ref{def:pi}, \ref{def:M}, \ref{def:delta}, and the semantics of FOL.
\end{proof}

\section{Connection method}

The connection method (CM)~\cite{bibel:cm} consists in a validity procedure (in opposition to refutation procedures such as tableaux and resolution), i.e., it tries to prove whether a formula (query) is valid directly. 
If $X=\{\varphi_{1},\ldots,\varphi_{n}\}$ is a (finite) set of first-order formulae (also referred to as a knowledge base, $\KB$), then in order to check whether $X\models\psi$, for some first-order formula~$\psi$, the validity of the formula $(\varphi_{1}\land\ldots\land\varphi_{n})\to\psi$ ($X\to\psi$), i.e., of $\nao{X}\lor\psi$, must be proven.
Hence, CM requires the conversion of formulae into the disjunctive normal form (DNF).
The consequences of negating the $\KB$ are the following: (\emph{i}) subsumption axioms (e.g., $E \sub D$) are converted into $E\land \nao{D}$; (\emph{ii}) variables are existentially quantified, instead of universally; (\emph{iii}) thus, FOL Skolemization applies over universally quantified variables, and (\emph{iv}) the consequent (or query) $\alpha$ is not negated.
Of course, when moving to the DL case, the crux of the matter is precisely how to express the negation of the knowledge base, along with doing away with variables and Skolem functions.

\alctcm\ \cite{fred:calculus} was the first CM proposed for DLs. Among its main features is the fact it requires neither variables in the initial representation of the axioms, nor Skolem functions. Moreover, it includes a blocking solution to ensure termination, as commonly done in the field, and the definition of $\theta$-substitution as a suitable replacement for variable unification.
Furthermore, a connection calculus for a defeasible extension of DLs has been proposed earlier~\cite{alcbtcm} for a less expressive language \alcb. It handles the preference relation as an auxiliary structure and relies on a normal form, the Typicality Normal Form.

When looking at the matrix, if we change our perspective, we can see the paths of the matrix.

\begin{definition}[Path]
Given a matrix \(M\), a \df{path} is a set containing exactly one literal from each clause of \(M\).
\end{definition}

If the matrix is the clausal form for a DNF formula, then its paths compose the clausal form for the same formula in CNF. Therefore, an interpretation satisfies a path iff the interpretation satisfies every literal on it. 
Furthermore, an interpretation satisfies a set of paths iff the interpretation satisfies at least one path.

\begin{lemma}
Given a matrix \(M\) and an interpretation \(\II\), \(\II\) satisfies \(M\) iff \(\II\) satisfies every path of it.
\end{lemma}


\begin{proof}
We prove this lemma by contrapositive on both sides.

\paragraph{Only-if part:} Assume that \(\II \not\sats p\), for some path \(p\) of \(M\). So, \(\II \not\sats L\), for every \(L \in p\). By the definition of a path, the path must contain a literal from each clause in the matrix, meaning that \(\II \not\sats C\), for every clause \(C \in M\). Therefore, \(\II \not\sats M\).

\paragraph{If-part:}. Assume that \(\II \not\sats M\). Then, at least one literal from each clause is not satisfied by \(\II\). By the semantics and the definition of a path, there exists a path containing each literal not satisfied by \(\II\) and, therefore, there exists a path not satisfied by \(\II\).
\qed\end{proof}

\begin{definition}[Multiplicity]
The \df{multiplicity} is a function \(\mu\) that maps for each clause of a matrix a natural number denoting the number of copies of such clause.
\end{definition}

We represent a matrix \(M\) and its copies as \(M^\mu\).

\begin{corollary}
Given a matrix \(M\), a multiplicity \(\mu\), and an interpretation \II, \(\II \sats M\) iff \(\II \sats M^\mu\).
\end{corollary} 

The proof idea is to show that copies preserve the semantics of a matrix. Due to the Skolemization process, there are only existentially quantified variables. Since a matrix is a formula in DNF, given an interpretation \(\II\):
\begin{equation*}
\II \sats \exists x_1, \cdots , x_n \varphi \text{~iff~} \exists x_1, \cdots , x_n~\varphi \lor \exists y_1, \cdots , y_n~\varphi[x_1 \mapsto y_1, \cdots, x_n \mapsto y_n]
\end{equation*}

\begin{definition}[Term substitution]
A \df{term substitution} is a function \(\sigma: \mathsf{V} \longrightarrow \mathsf{T}\) that maps variables to terms of a matrix. We define that \(\sigma(L)\) is the literal \(L\), but its variables \(x\) are replaced by \(\sigma(x)\), i.e., \(\sigma(A(x)) = A(\sigma(x))\).
\end{definition}

Besides this definition, we shall ensure no substitution between a variable and a term that mentions it. A substitution with this behaviour is called \df{idempotent}. However, we will omit this term hereafter to a better understanding.

\begin{definition}[\(\sigma\)-complementary literals]
Given a literal \(L\), we say that \(\overline{L}\) is the \df{$\sigma$-complement of} \(L\) if \(\sigma(\overline{L}) = \sigma(\neg P)\), when \(L = P\) or \(\sigma(\overline{L}) = \sigma(P)\) when \(L = \neg P\).
\end{definition}

Hereafter, in order to decrease the complexity of definitions, we shall omit \(\sigma\) before complementary literals or sets.

\begin{definition}[Complementary set]
A \df{complementary set} is a pair \(\{L_1, L_2\}\), where \(\sigma(L_1) = \sigma(\overline{L_2})\).
\end{definition}

\begin{definition}[Complementary matrix]
A matrix is called \df{complementary} if there exists a term substitution \(\sigma\) such that it is a complementary set for each path on it.
\end{definition}

\begin{theorem}[Matrix characterisation] \label{theorem:matrix.characterisation}
Given a matrix \(M\), there exists a multiplicity \(\mu\) and a term substitution \(\sigma\) such that \(\sigma(M^\mu)\) is complementary iff \(\models M\).
\end{theorem}

\begin{proof}
We prove this theorem by contrapositive.

\(\Rightarrow\). Assume that \(\not\models M\). Then, there exists an interpretation \(\II\) such that \(\II \not \sats M\). Hence, there exists a path in \(M\) that is not satisfied by \(\II\) since the path corresponds to a disjunction in CNF. That path cannot contain a complementary set on it, no matter which substitution is applied to it. Otherwise, it is a validity, and \(\II\) must satisfy it. Therefore, \(M\) is not complementary under any term substitution.

\(\Leftarrow\). Assume that \(\sigma(M^\mu)\) is not complementary, for all multiplicity \(\mu\) and for all term substitution \(\sigma\). Then, there exists a path which does not contain complementary sets on it. Hence, we can build an interpretation such that it does not satisfy every literal of the path. Let the path be \(p\), we build an interpretation \(\II\) such that:
\begin{itemize}
\item \(\II \sats \neg{A}(t)\), for all unary predicate \(A\) and term \(t\) s.t. \(A(t) \in p\);
\item \(\II \sats A(t)\), for all unary predicate \(A\) and term \(t\) s.t. \(\neg{A}(t) \in p\);
\item \(\II \sats \neg{R}(t, u)\), for all binary predicate \(R\) and terms \(t, u\) s.t. \(R(t, u) \in p\);
\item \(\II \sats R(t, u)\), for all binary predicate \(R\) and terms \(t, u\) s.t. \(\neg{R}(t, u) \in p\);
\item \(\II \sats \neg{(t < u)}\), for all terms \(t, u\) s.t. \((t < u) \in p\);
\item \(\II \sats (t < u)\), for all terms \(t, u\) s.t. \(\neg{(t < u)} \in p\);
\item \(\II \sats \neg{((t, u) \ll (v, k))}\), for all terms \(t, u, v, k\) s.t. \(((t, u) \ll (v, k)) \in p\); and
\item \(\II \sats ((t, u) \ll (v, k))\), for all terms \(t, u, v, k\) s.t. \(\neg{((t, u) \ll (v, k))} \in p\).
\end{itemize} 
That interpretation does not satisfy the formula in CNF. Otherwise, it should satisfy the path. Therefore, \(\not\models M\).
\qed\end{proof}

\section{Connection calculus}

We define the connection method with a formal calculus. The main structure of the calculus is a triple \(\langle C, M, \mathit{Path} \rangle\), where \(C\) is the goal, \(M\) is the matrix, and \textit{Path} is the active path of the proof. The proof starts with no goal and path (represented by \(\varepsilon\)), applying the \df{Start rule} for some clause in \(M\). If the proof finds an empty set as the goal, then that branch is closed with the \df{Axiom}. 

As \alchb\ is a decidable-fragment of preferential DLs, a notion of blocking is needed before presenting the calculus.

\begin{definition}[Set of concepts]
Given a term \(t\), a path \(p\), and a term substitution \(\sigma\), the \df{set of concepts} of \(t\) w.r.t. the path \(p\), denoted by \(\tau^\sigma_p(t)\) is
\[
\tau^\sigma_p(t) = \{D \mid \text{for all } D(\sigma(t)) \in \sigma(p)\} 
\]
\end{definition}

\begin{definition}[Copy]
Given two clauses \(C_1\) and \(C_2\), we say that \(C_2\) is a \df{copy} of \(C_1\) if they have the same literals but different variables.
\end{definition}

\begin{definition}[\(i\)-th copy]
Given a matrix \(M\), a clause \(C \in M\), and a multiplicity \(\mu\), with \(C^i\) we denote the \df{$i$-th copy} of \(C \in M^\mu\), where \(1 \leq i \leq \mu(C)\).
\end{definition}

\begin{definition}[Blocking]
Given a literal \(L\) and a path \(p\), we say that \(p \cup \{L\}\) is \df{blocked} w.r.t. a term substitution \(\sigma\) if:
\begin{enumerate}
\item \(\sigma(L) \in \sigma(p)\); or
\item if \(L \in C^n\), for some clause \(C \in M^\mu\) and some \(2 \leq n \leq \mu(C)\), there exists a previous copy term \(t_{n-1}\) such that either \(t_{n-1}\) is blocked or \(\tau_{p \cup \{L\}}^\sigma(t_{n}) \subseteq \tau_{p \cup \{L\}}^\sigma(t_{n-1})\).
\end{enumerate}
\end{definition}

\begin{definition}[Blocked cause]
We say that a \df{clause is blocked} w.r.t. a path \(p\) and a term substitution \(\sigma\) if some literal \(L \in C\) is blocked w.r.t. \(p\).
\end{definition}

During the proof, the calculus connects literals on the goal in two different ways. The first one, \df{Reduction rule}, occurs when a complementary literal is already on the active path, meaning that path contains a complementary set. The second one, \df{Extension rule}, occurs when it finds another clause with a complementary literal for the goal. In this case, the proof is branched into two parts in order to check the remaining goal (without that literal complemented) and the new goal (the clause found).
Every application of a rule in calculus must ensure the goal is not blocked w.r.t. to its path. Otherwise, no rule application must occur.

\begin{figure}
    \begin{equation*}
        \text{\textbf{\textit{Axiom (Ax)}}} ~ \dfrac{}{\{\}, M, \mathit{Path}}
    \end{equation*}
    
    \begin{equation*}
        \text{\textbf{\textit{Start rule (St)}}} ~  \dfrac{C_1,M,\{\}}{\varepsilon,M,\varepsilon}, \ \text{with } C_1 \text{ being a blocking-free copy of some } C \in M
    \end{equation*}
    
    \begin{equation*}
    \begin{gathered}
        \text{\textbf{\textit{Reduction rule (Red)}}} ~ \dfrac{C,M,\mathit{Path} \cup \{L_2\}}{C \cup \{L_1\},M,\mathit{Path} \cup \{L_2\}}, 
        \\ \text{with } \sigma(L_1) = \sigma(\overline{L_2})
    \end{gathered}
    \end{equation*}
    
    \begin{equation*}
    \begin{gathered}
        \text{\textbf{\textit{Extension rule (Ext)}}} ~ \dfrac{C_1 \setminus \{L_2\},M,\mathit{Path} \cup \{L_1\} ~ ~ ~ ~ C,M,\mathit{Path}}{C \cup \{L_1\},M,\mathit{Path}}, \\
        \text{with } L_2 \in C_1, C_1 \text{ being a blocking-free copy of some } C_2 \in M, \text{ and } \sigma(L_1) = \sigma(\overline{L_2})
    \end{gathered}
    \end{equation*}
    

    \caption{The calculus.}
    \label{fig:calculus}
\end{figure}

\begin{definition}[Connection proof]
Given a triple \(\langle C, M, \textit{Path} \rangle\), we say that it is a \df{connection proof} if, applying the rules of the calculus for \(\langle C, M, \textit{Path} \rangle\), there exists a multiplicity \(\mu\), a substitution \(\sigma\), and a proof tree such that every leaf ends with an \df{Axiom}.
\end{definition}

\begin{definition}[Relative paths]
Given two sets of literals \(C\) and \(S\), and a matrix \(M\), we define the \df{relative paths} of \(C\), \(S\)  and \(M\) as:
\begin{align*}
\varphi(M) &\defined \{p \mid p \text{ is a path of }M\} \\
\varphi(M, S) &\defined \{p \in \varphi(M) \mid S \subseteq p\} \\
\varphi(M, S, C) &\defined \{p \in \varphi(M, S) \mid L \in p, \text{for some } L \in C \}
\end{align*}
\end{definition}

\begin{corollary}
\begin{equation*}
\varphi(M, S \cup \{L\}) = \varphi(M, S, \{L\})
\end{equation*}
\end{corollary}

\begin{proposition} \label{prop:soundness}
Given a triple \(\langle C, M, \textit{Path} \rangle\), if it is a connection proof, for some variable substitution \(\sigma\), then there exists a multiplicity \(\mu\) for all path \(p \in \varphi(M, \text{Path}, C)\), s.t. \(p\) is \(\sigma\)-complementary.
\end{proposition}

\begin{proof}
    We prove the lemma above by structural induction over the connection proofs.
    Since a connection proof can be a subtree of another connection proof, we can assume, as the Induction Hypothesis (IH), that if there exists a connection proof of a subtree for some \(\sigma\), then there exists a multiplicity \(\mu\) such that every path in it is \(\sigma\)-complementary.
    We demonstrate that the lemma holds for Axiom and the rules Reduction and Extension in the calculus as follows:
    
    \begin{itemize}
        \item \textbf{Axiom (Ax) :} Assume that \AxiomC{}\UnaryInfC{$\{\}, M, \textit{Path}$}\DisplayProof is a connection proof for $\langle \{\}, M, \textit{Path} \rangle$. Therefore, $\varphi(M,\textit{Path},\emptyset) = \emptyset$, meaning that IH holds since there is no complementary path in the empty set;
            
        \item \textbf{Reduction (Red) :} Assume that \AxiomC{\ensuremath{\mathit{Proof}}}\UnaryInfC{$C, M, \textit{Path} \cup \{L_2\}$}\DisplayProof is a connection proof for $\langle C, M, \textit{Path} \cup \{L_2\} \rangle$, for some variable substitution \(\sigma\).
        Then, the derivation
        \AxiomC{\ensuremath{\mathit{Proof}}}\UnaryInfC{$C, M, \textit{Path} \cup \{L_2\}$}\UnaryInfC{$C \cup \{L_1\}, M, \textit{Path} \cup \{L_2\}$}\DisplayProof
        is a connection proof, where $\sigma'(\sigma(L_1)) = \sigma'(\sigma(\overline{L_2}))$.
        By the IH, for some multiplicity \(\mu\), every path $p' \in \varphi(M^\mu, \textit{Path} \cup \{L_2\}, C)$ is \(\sigma\)-complementary.
        Furthermore, every path $p'' \in \varphi(M^\mu, \textit{Path} \cup \{L_2\}, \{L_1\})$ is \(\sigma'\)-complementary, since $\sigma'(\sigma(L_1)) = \sigma'(\sigma(\overline{L_2}))$.
        Let \(\mu'\) be \(\mu\) and \(\sigma''\) be the composition of \(\sigma\) and \(\sigma'\).
        As $\varphi(M^{\mu'},\textit{Path} \cup \{L_2\},C\cup\{L_1\}) = \varphi(M^{\mu'}, \textit{Path} \cup \{L_2\}, C) \cup \varphi(M^{\mu'}, \textit{Path} \cup \{L_2\}, \{L_1\})$, we conclude that every path in $\varphi(M^{\mu'}, \textit{Path}\cup\{L_2\}, C\cup\{L_1\})$ is \(\sigma''\)complementary;

        \item \textbf{Extension (Ex) :} Assume that \AxiomC{\ensuremath{\mathit{Proof1}}}\UnaryInfC{$C_2 \setminus \{L_2\}, M, \textit{Path} \cup \{L_1\}$}\DisplayProof is a connection proof for $\langle C_2 \setminus \{L_2\}, M, \textit{Path} \cup \{L_1\} \rangle$, and \AxiomC{\ensuremath{\mathit{Proof2}}}\UnaryInfC{$ C, M, \textit{Path}$}\DisplayProof is a connection proof for $\langle C, M, \textit{Path} \rangle$, for some substitution \(\sigma\).
        By the IH, there exists a multiplicity \(\mu_1\) such that every path in \(\varphi(M^{\mu_1}, \textit{Path} \cup \{L_1\}, C_2 \setminus \{L_2\})\) is \(\sigma\)-complementary, and there exists a multiplicity \(\mu_2\) such that every path in \(\varphi(M^{\mu_2}, \textit{Path}, C)\) is \(\sigma\)-complementary.
        Then, the derivation
            \begin{center}
                \AxiomC{Proof1}
                \UnaryInfC{$C_2 \setminus \{L_2\}, M, \textit{Path} \cup \{L_1\}$}
                \AxiomC{Proof2}
                \UnaryInfC{$C, M, \textit{Path}$}
                \BinaryInfC{$C \cup \{L_1\},M,\textit{Path}$}
                \DisplayProof
            \end{center}
        with $\sigma'(\sigma(L_1)) = \sigma'(\sigma(\overline{L_2}))$ and $C_2$ as a copy of some clause $C_1 \in M$, is a connection proof for $\langle C \cup \{L_1\},M,\textit{Path} \rangle$.
        Let \(\mu'\) be the combination of \(\mu_1\) and \(\mu_2\), and \(\sigma''\) be the composition of \(\sigma\) and \(\sigma'\).
        Hence, every path $p' \in \varphi(M^{\mu'}, \textit{Path}, C)$ is \(\sigma''\)-complementary and every path $p'' \in \varphi(M^{\mu'}, \textit{Path}, C_2 \setminus \{L_2\})$ is also \(\sigma''\)complementary.
        Moreover, once $\sigma''(L_1) = \sigma''(\overline{L_2})$, every path in $\varphi(M^{\mu'}, \textit{Path}\cup\{L_1\}, \{L_2\})$ is complementary.
        As $C_2 \in M^{\mu'}$, we also have that $\varphi(M^{\mu'}, \textit{Path}\cup\{L_1\}, C_2) = \varphi(M, \textit{Path}\cup\{L_1\})$ and every path on it is complementary.
        Therefore, every path in $\varphi(M^{\mu'}, \textit{Path}, C \cup \{L_1\})$ is \(\sigma''\)-complementary since $\varphi(M^{\mu'}, \textit{Path}, C \cup \{L_1\}) = \varphi(M^{\mu'}, \textit{Path} \cup \{L_1\}) \cup \varphi(M, \textit{Path}, C)$.
    \end{itemize}
\qed\end{proof}

\begin{lemma}[Soundness of the calculus] \label{lemma:soundness}
Given a matrix \(M\), if \(\langle \varepsilon, M, \varepsilon \rangle\) is a connection proof with \(\sigma\), then there exists a multiplicity \(\mu\) such that every path in \(\varphi(M^\mu)\) is \(\sigma\)-complementary.
\end{lemma}

\begin{proof}
We prove this theorem by contrapositive. If we assume there is no multiplicity, then no copies are able to occur, and the triple is not a connection proof. Now, we assume there exists a multiplicity \(\mu\) and a path in \(\varphi(M^\mu)\) that is not \(\sigma\)-complementary, for all \(\sigma\). 
Let 
    \AxiomC{$\cdots$}
    \UnaryInfC{$\langle C_2, M, \{\} \rangle$}
    \RightLabel{\textit{St}}
    \UnaryInfC{$\langle \varepsilon ,M, \varepsilon \rangle$}
    \DisplayProof 
be the derivation for $M$, with $C_2$ being a copy of some clause $C_1 \in M$.
Then, \(\varphi(M^\mu, \{\}, C_2)\), is not \(\sigma\)-complementary as well. By Proposition \ref{prop:soundness}'s contrapositive, there is no connection proof for \(\langle C_2, M, \{\} \rangle\), therefore, there is no connection proof for \(\langle \varepsilon, M, \varepsilon\).
\qed\end{proof}

\begin{lemma}[Completeness of the calculus] \label{lemma:completeness}
Given a matrix \(M\), if there exists a multiplicity \(\mu\) and variable substitution \(\sigma\) such that every path in \(\varphi(M^\mu)\) is \(\sigma\)-complementary, then \(\langle \varepsilon, M, \varepsilon \rangle\) is a connection proof.
\end{lemma}

\begin{proof}
So, we prove this theorem by contrapositive. Then, if there is no connection proof for $\langle \epsilon, M, \epsilon \rangle$, then there exists no multiplicity \(\mu\) and variable substitution \(\sigma\) such that every path in \(\varphi(M^\mu)\) is \(\sigma\)-complementary.
    Thus, w.l.o.g. there exists a satured derivation branch of $\langle \epsilon, M, \epsilon, \rangle$ containing a leaf 
        \AxiomC{$\langle C, M, \mathit{Path} \rangle$}
        \UnaryInfC{$\cdots$}
        \DisplayProof 
    such that there is no rule of the calculus to be applied. 
    The clause $C$ cannot be empty. Otherwise, the Axiom would be applied. Also, the clause $C$ does not contain a literal $L$ such that its complement $\overline{L} \in \mathit{Path}$ with \(\sigma\). Otherwise, the Reduction rule would be applied. 
    Finally, there is no blocking-free copy clause $C_2$ of \(C_1 \in M^\mu\) such that $\overline{L} \in C_2$, for some literal \(L \in C\), otherwise the extension rule would be applied.
    To enforce this possibility, assume there exists a clause blocked by \(\mathit{Path}\), and it contains a complementary literal. Then, once it is blocked, it means that: (i) it exists in the \(\textit{Path}\), so the path would remain the same, or (ii) there exists a previous copy term such that its set of concepts  is the same or greater than the set of concepts of the target term in the new clause. Once this derivation is a saturated proof, this clause was used before in the path, and it remains open. It means there exists an infinite loop for this proof, and as such, it is blocked, remaining open to this derivation.
    Therefore, there exists a path $p \in \varphi(M^\mu,\mathit{Path})$ over the matrix $M$ such that $p$ is not complementary for any multiplicity \(\mu\).
\qed\end{proof}

\begin{algorithm}[H]
\caption{Inconsistent}\label{alg:inconsistent}
\Inconsistent{$(\K)$}{
\Input{An \alchb\ knowledge base $\K$}
\Output{\textit{True} if $\K \models \bot$ or \textit{False} otherwise}
    $F \gets \delta(\K)$\;
    $M$ stores the DNF clausal form of formula \(F\) after Herbrandization \;
    \For{each clause $C \in M$}{
        \If{\textbf{Proof}$(C, M, \{\})$ is True}{
            \KwRet{True}\;
        }
    }
    \KwRet{False}\;
}
\end{algorithm}

\begin{algorithm}[H]
\caption{Proof}\label{alg:proof}
\Proof{$(C, M, \mathit{Path})$}{
\Input{A (sub-)clause $C$, a matrix $M$ and a (sub-)path $\mathit{Path}$}
\Output{\textit{True} if there exists a connection proof for $\langle C, M, \mathit{Path} \rangle$ or \textit{False} otherwise}
    \Comment{ for Axiom}
    \If{$C = \emptyset$}{
        \KwRet{True}\;
    }
    \For{each rule $R$ that is applicable to $\langle C, M, \mathit{Path} \rangle$}{
        \For{each triple \(\langle C', M', \textit{Path}' \rangle\) derived from applying $R$}{
            \If{\textbf{Proof}\((C', M', \textit{Path}')\) is False}{
                skip rule $R$\;
            }
        }
        \KwRet{True}\;
    }
    \KwRet{False}\;
}
\end{algorithm}

\begin{theorem}[Termination]
   Given any \alchb\ knowledge base $\K$, \df{\textbf{Inconsistent}}($\K$) terminates.
\end{theorem}

\begin{proof}
The key component that affects the termination of the algorithm is when Extension rules are applied. The reduction rule reduces the number of literals to be proven, and Axiom closes the branch with an empty goal. Hence, they lead to a finite number of applications.
The Extension rule reduces the number of literals to be proven on one side of its derivation but adds a new clause to be proven on its left side.
So, to ensure the application of Extension rules stops, we need to prove that at some point, every clause with complementary literals of a goal is blocked.
	
Let \(m\) be the number of concept names occurring in \(\K\). During an Ext application, there exists at least a term \(t\) in the new clause that is unified with a term \(u\) in the path. Once there exists a finite number of concept names, the term \(t\) can appear at most \(m\) times in the path before being blocked (Case 1 of Blocking definition). The same idea can be used for pair of terms since there exists at most \(n\) role names.
If the Ext application adds new terms to the path, it must add different concept names to the path to prevent being blocked by its previous copy term. Again, there exists a finite number of concept names. Therefore, the blocking will occur at some point.
	
%
\qed\end{proof}

\begin{theorem}[Soundness]
Given a knowledge base \(\K\), if \df{\textbf{Inconsistent}}\((\K)\) returns \(\mathit{True}\), then \(\K\) is inconsistent.
\end{theorem}

\begin{proof}
The proof is a direct consequence of Theorem \ref{theorem:matrix.characterisation} and Lemma \ref{lemma:soundness}.
\qed\end{proof}

\begin{theorem}[Completeness]
Given a knowledge base \(\K\), if \(\K\) is inconsistent, then \df{\textbf{Inconsistent}}(\(\K\)) returns \(\mathit{True}\)
\end{theorem}

\begin{proof}
The proof is a consequence of Theorem \ref{theorem:matrix.characterisation} and Lemma \ref{lemma:completeness}.
\qed\end{proof}

\section{Example of proof}\label{Section:ExampleOfProof}

We present in this section an example to illustrate how the proposal checks whether a knowledge base entails an assertion. For that, assume \(\CN = \{A, B\}\), \(\RN = \{r, s\}\), \(\IN = \{a, b\}\) as the set of concept names, set of role names, and set of individual names, respectively.
Let \(\K = (\T, \R, \A)\) be a knowledge base, where \(\T = \{\exists s . B \sub \pf A\}\), \(\R = \{\pf r \sub s\}\), and \(\A = \{\pf r(a, b), B(b)\}\). We want to check whether \(\K \models A(a)\), i.e., that \(A(a)\) is entailed by \(\K\). 
The first step is to translate \(\K\) and \(A(a)\) to the matrix \(M = \delta(\K) \cup \varrho(A, a, \varepsilon)\) as

\begin{align*}
&\varrho(\exists s . B \e \nao{\pf A}, x_1, x_1)) 
\cup \varrho(\pf r, x_2, y_1) \clup \varrho(\nao{s}, x_2, y_1) \\
&\cup \varrho(\nao{\pf r}(a, b))
\cup \varrho(\nao{B}(b, \varepsilon))
\cup \varrho(A, a, \varepsilon)
\cup \Phi
\end{align*}

and its graphical representation is shown in Figure \ref{fig:matrix}.

\begin{figure}
\resizebox{1.2\textwidth}{!}{
\(
\cmatrix{
s(x,y) \& B(y) \& \nao{A}(x) \\
s(x,y) \& B(y) \& x' < x \& A(x') \\
r(x, y) \& \nao{((f(x), g(y)) \ll (x, y))} \& \nao{s}(x, y) \\
r(x, y) \& \nao{r}(f(x), g(y)) \& \nao{s}(x, y) \\
r(x, y) \& (x', y') \ll (f(x), g(y)) \& r(x', y') \& \nao{s}(x, y) \\
\nao{r}(a, b) \\
(x', y') \ll (a, b) \& r(x', y') \\
\nao{B(b)} \\
A(a) \\
\Phi \\
}{}
\)
}
\caption{Matrix representation of \(\delta(\K) \cup \varrho(A, a, \varepsilon)\). The matrix displays clauses as columns in the same order as shown in $\K$.}
\label{fig:matrix}
\end{figure}

It may not be necessary to use all the clauses of the matrix besides the connection method making connections through all the paths. 
In our example, the second clause and the transitivity, irreflexivity, and asymmetry representation are not used.\footnote{As mentioned in Section~\ref{TheMethod}, axioms are denoted by~\(\Phi\) in the matrix.} We shall therefore omit them from now on.

\begin{figure}
\centering
\resizebox{1.2\textwidth}{!}{
\(
\cmatrix{
A(a) \\
s(x_1, y_1) \& B(y_1) \& \nao{A}(x_1) \\
\nao{B(b)} \\
\nao{s}(x_2, y_2) \& r(x_2, y_2) \& \nao{r}(f(x_2), g(y_2)) \\
\nao{r}(a, b) \\
(x_3, y_3) \ll (a, b) \& r(x_3, y_3) \\
\nao{s}(x_4, y_4) \& r(x_4, y_4) \& \nao{((f(x_4), g(y_4)) \ll (x_4, y_4))} \\
}{
\draw[solid, red] (m-1-1.south) to[bend left=20] node[midway,above] {\reflectbox{\rotatebox{270}{1$^{\{x_1/a\}}$}}} (m-2-3.west);
\draw[solid, red] (m-2-2.east) to[bend left=20] node[midway,below] {\reflectbox{\rotatebox{270}{2$^{\{y_1/b\}}$}}} (m-3-1.south);
\draw[solid, red] (m-2-1.north) to[bend right=20] node[pos=0.3, above] {\reflectbox{\rotatebox{270}{3$^{\{x_2/a, y_2/b\}}$}}} (m-4-1.north);
\draw[solid, red] (m-4-2.east) to[bend left=20] node[xshift=-0.7em, above=0] {\reflectbox{\rotatebox{270}{4}}} (m-5-1.south);
\draw[solid, red] (m-4-3.east) to[bend left=20] node[xshift=1.0em, yshift=0.5em] {\reflectbox{\rotatebox{270}{5$^{\{x_3/f(a), y_3/g(a)\}}$}}} (m-6-2.south);
\draw[solid, red] (m-6-1.east) to[bend left=5] node[xshift=1.0em, yshift=2.5em] {\reflectbox{\rotatebox{270}{6$^{\{x_4/a, y_4/b\}}$}}} (m-7-3.north);
\draw[solid, red] (m-7-2.west) to[bend left=25] node[xshift=-0.7em, yshift=0.5em] {\reflectbox{\rotatebox{270}{7}}} (m-5-1.north);
\draw[solid, red] (m-7-1.north) to[bend left=15] node[xshift=-0.7em, yshift=0.5em] {\reflectbox{\rotatebox{270}{8}}} (m-2-1.north);
}
\)
}
\(\sigma = \{x_1 / a, y_1 / b, x_2 / a, y_2 / b, x_3 / f(x_2), y_3 / g(y_2), x_4 / a, y_4 / b\}\)
\caption{Connection method applied over \(M\). Non-used clauses are omitted for the sake of space.}
\label{fig:matrix.connection}
\end{figure}

In Figure \ref{fig:matrix.connection}, we perform the connection method in a graphical way. Red curves between literals of the matrix represent the application of the Start, Extension, or Reduction rules. The matrix was rearranged to avoid misunderstanding when connecting. The goal is to check if every clause associated with a connection is fully connected. Otherwise, the derivation is not a connection proof.
It starts with \(A(a)\), the possible entailed assertion, connecting it to a complementary literal \(\nao{A}(a)\) in the second clause. As there is no \(\nao{A}(a)\) in the matrix, we perform a substitution to \(x_1\) to make \(\sigma\)-substitution and unify the literals. Therefore, \(\sigma = \{x_1/a\}\) at this point. 
Now, the second clause is our subgoal, and the other two literals have to be proven. We connect \(B(y_1)\) with \(\nao{B(b)}\) in the third clause, where \(\sigma(y_1) = b\). As there is no other literal in the third clause, we go back to the second one, performing a connection between \(s(x_1, y_1)\) and \(\nao{s}(x_2, y_2)\) in the fourth clause.
The reader may notice that we can connect the \(s(x_1, y_1)\) with \(\nao{s}(x_4, y_4)\) in the last clause. In fact, the connection would head the proof to the same conclusion.
The other connections follow the same approach, but we want to focus on the last connection (8). As we have the \(\sigma\)-complementary literal in the active path, this connection is a reduction instead of the extension rule application for the other ones. Therefore, as every clause is fully connected, we prove that \(\K \models A(a)\).

Another useful representation of the connection method is the sequent-style derivation tree. It shows the triples during the search proof and applies the rules until no rule can be applied. If every leaf is an Axiom, i.e., the subgoal is fully proven, then it is a connection proof of \(M\). Figure \ref{fig:example.calculus} shows this representation.

\begin{figure}

\centering

\resizebox{.7\textwidth}{!}{
\def\defaultHypSeparation{\hskip.05in}
\def\ScoreOverhang{0pt}
\def\labelSpacing{4pt}
\AxiomC{}
\RightLabel{Ax}
\UnaryInfC{$\{\}, M, \mathit{Path_4}$}
\AxiomC{}
\RightLabel{Ax}
\UnaryInfC{$\{\}, M, \mathit{Path_3}$}
\RightLabel{Ext}
\BinaryInfC{$\{r(x_4, y_4)\}, M, \mathit{Path_3}$}
\RightLabel{Red}
\UnaryInfC{$\{\nao{s(x_4, y_4)}, r(x_4, y_4)\}, M, \mathit{Path_3}$}
\AxiomC{}
\RightLabel{Ax}
\UnaryInfC{$\{\}, M, \mathit{Path_2}\}$}
\RightLabel{Ext}
\BinaryInfC{$\{(x_3, y_3) \ll (a, b)\}, M, \mathit{Path_2}$}
\AxiomC{}
\RightLabel{Ax}
\UnaryInfC{$\{\}, M, \mathit{Path_1}$}
\RightLabel{Ext}
\BinaryInfC{$\{\nao{r(f(x_2), g(y_2))}\}, M, \mathit{Path_1}$}
\UnaryInfC{\textbf{\textit{Proof}}}
\DisplayProof
}

\vspace{0.5em}

\resizebox{\textwidth}{!}
{
\def\defaultHypSeparation{\hskip.05in}
\def\ScoreOverhang{0pt}
\def\labelSpacing{4pt}
\AxiomC{}
\RightLabel{Ax}
\UnaryInfC{$\{\}, M, \{A(a), s(x_1, y_1), r(x_2, y_2)\}$}
\AxiomC{\textbf{\textit{Proof}}}
\RightLabel{Ext}
\BinaryInfC{$\{r(x_2, y_2), \nao{r(f(x_2), g(y_2))}\}, M, \mathit{Path_1}$}
\AxiomC{}
\RightLabel{Ax}
\UnaryInfC{$\{\}, M, \{A(a), B(y_1)\}$}
\RightLabel{Ext}
\AxiomC{}
\RightLabel{Ax}
\UnaryInfC{$\{\}, M, \{A(a)\}$}
\RightLabel{Ext}
\BinaryInfC{$\{B(y_1)\}, M, \{A(a)\}$}
\RightLabel{Ext}
\BinaryInfC{$\{s(x_1, y_1), B(y_1)\}, M, \{A(a)\}$}
\AxiomC{}
\RightLabel{Ax}
\UnaryInfC{$\{\}, M \{\}$}
\RightLabel{Ext}
\BinaryInfC{$\{A(a)\}, M, \{\}$}
\RightLabel{\textit{Start}}
\UnaryInfC{$\varepsilon, M, \varepsilon$}
\DisplayProof
}
\begin{small}
\begin{align*}
\mathit{Path_1} &= \{A(a), s(x_1, y_1)\} \\
\mathit{Path_2} &= \{A(a), s(x_1, y_1), \nao{r(f(x_2), g(y_2))}\} \\
\mathit{Path_3} &= \{A(a), s(x_1, y_1), \nao{r(f(x_2), g(y_2))}, (x_3, y_3) \ll (a, b)\} \\
\mathit{Path_4} &= \{A(a), s(x_1, y_1), \nao{r(f(x_2), g(y_2))}, (x_3, y_3) \ll (a, b), r(x_4, y_4)\}
\end{align*}
\end{small}
\label{fig:example.calculus}

\caption{The connection method in sequent-style for \(M\). It starts from the bottom to the top. As every leaf is an Axiom, we call this proof derivation a connection proof.}
\end{figure}


\section{Concluding remarks}\label{Conclusion}

In this work, we have defined a connection method for \alchb, a defeasible description logic at least as expressive as most of the defeasible DLs considered in the literature. The calculus extends \alctcm, in two aspects: (\emph{i})~it complies with the preferential-DL semantics developed by Britz et al.\ and by Giordano et al.\ , which is widely assumed in the literature on reasoning with defeasible ontologies; and (\emph{ii})~it relies on a tailor-made matrix translation we introduced to cater for typicality in concepts and in roles.

As already alluded to in the introduction, the work reported here is part of a pioneer study of connection methods as viable alternatives for reasoning with defeasible ontologies. It is, therefore, part of a broader long-term agenda, of which an immediate next step is endowing \textsc{Raccoon} with the ability to reason over defeasible extensions of~\alch.

The reader conversant with preferential reasoning would have noticed that in this work we assume preferential entailment, which is a Tarskian notion of consequence and, therefore, monotonic. As pointed out in the literature on non-monotonic reasoning, preferential entailment is not always enough for reasoning defeasibly with exceptions. 
Stronger, more venturous forms of entailment are often called for. One particular definition thereof, namely the \textit{rational closure} of a defeasible ontology, has been thoroughly investigated in the context of defeasible \alc~\cite{BritzEtAl2021,CasiniStraccia2010,GiordanoEtAl2015}. Nevertheless, a case has been made for sticking to preferential reasoning in some contexts~\cite{GiordanoEtAl2010} or for investigating weaker forms of rationality~\cite{Freund1993}. This suggests the debate around rational closure being the baseline for defeasible reasoning remains, in a sense, open.

In any case, the aforementioned approaches to the computation of the rational closure of a defeasible knowledge base rely on a number of calls to a classical (monotonic) reasoner, and therefore the availability of a method for preferential reasoning in \alchb\ as the one we propose here is an important step in the investigation of stronger forms of entailment in the presence of typicality operators.

The limitations of adopting a single preference ordering in modelling object typicality have already been pointed out by Britz and Varzinczak~\cite{BritzVarzinczak2019-TABLEAUX}. They introduce a notion of context in multi-preference semantics, making it possible for some objects to be more typical than others w.r.t.\ a context but less typical w.r.t.\ a different one. Part of our research agenda is, therefore, to extend the method proposed here to deal with multiple preference relations (on both sets of objects and pairs of objects).


\section*{Acknowledgments}

This study has been financed in part by the Coordenação de Aperfeiçoamento de Pessoal de Nível Superior - Brasil (CAPES) - Finance Code 001. It also has been supported in part by the project \df{Reconciling Description Logics and Non-Monotonic Reasoning in the Legal Domain} (PRC CNRS–FACEPE, France–Brazil). Fred Freitas is partially supported by a Research Productivity grant provided by the Brazilian funding agency  National Council for Scientific and Technological Development – CNPq.

This work was partially supported by the ANR Chaire IA BE4musIA: BElief change FOR better MUlti-Source Information Analysis.


\bibliographystyle{splncs04}
\bibliography{biblio}

\end{document}